\documentclass[preprint,12pt]{elsarticle}

\makeatletter
\def\ps@pprintTitle{%
  \let\@oddhead\@empty
  \let\@evenhead\@empty
  \def\@oddfoot{}%
  \let\@evenfoot\@oddfoot}
\makeatother

\usepackage{amssymb}
\usepackage{amsmath}
\usepackage{booktabs}

\usepackage{adjustbox}
\usepackage[table]{xcolor}
\usepackage{colortbl}
\usepackage{subfig}         
\usepackage[para,online,flushleft]{threeparttable}
\usepackage[most]{tcolorbox}
\usepackage{hyperref}
\usepackage{multirow}

\definecolor{col1}{RGB}{240, 248, 255}  
\definecolor{col2}{RGB}{245, 255, 250}  
\definecolor{col3}{RGB}{255, 250, 240}  
\definecolor{col4}{RGB}{255, 245, 238}  
\definecolor{col5}{RGB}{250, 240, 255}  
\definecolor{rowalt}{gray}{0.95}        

\newcommand{\bq}{\begin{equation}}
\newcommand{\eq}{\end{equation}}

\newcommand{\second}{\mbox{s}}

\newcommand{\bit}{\mbox{bit}}

\newcommand{\GBPS}{\mbox{G\bit/\second}}

\newcommand{\GBS}{\mbox{GB/\second}}

\newcommand{\GHZ}{\mbox{GHz}}

\newcommand{\GiB}{\mbox{GiB}}
\newcommand{\MiB}{\mbox{MiB}}
\newcommand{\KiB}{\mbox{KiB}}

\newcommand{\eos}{~.}
\newcommand{\cma}{~,}

\colorlet{backgroundcol}{cyan!10!white}
\newcommand{\highlight}[1]{%
	\par\noindent
	\fcolorbox{black}{backgroundcol}{%
		\parbox{\dimexpr\linewidth-2\fboxsep\relax}{%
			#1
		}%
}}

\journal{}

\begin{document}

\begin{frontmatter}



\title{Exploring metrics for analyzing dynamic behavior in MPI programs via a coupled-oscillator model}


\author[NHR]{Ayesha Afzal\corref{cor1}}
\ead{ayesha.afzal@fau.de}
\ead[url]{https://orcid.org/0000-0001-5061-0438}
\cortext[cor1]{Corresponding author}

\author[NHR]{Georg Hager}
\ead{georg.hager@fau.de}
\ead[url]{https://orcid.org/0000-0002-8723-2781}

\author[CS]{Gerhard Wellein}
\ead{gerhard.wellein@fau.de}
\ead[url]{https://orcid.org/0000-0001-7371-3026}

\affiliation[NHR]{organization={Erlangen National High Performance Computing Center (NHR@FAU)},
            addressline={}, 
            city={Erlangen},
            postcode={91058},
            state={},
            country={Germany}}

\affiliation[CS]{organization={Department of Computer Science, Friedrich-Alexander-Universität},
            addressline={}, 
            city={Erlangen-Nürnberg},
            postcode={91058},
            state={},
            country={Germany}}

\begin{abstract}
We propose a novel, lightweight, and physically inspired approach to modeling the dynamics of parallel distributed-memory programs. Inspired by the Kuramoto model, we represent MPI processes as coupled oscillators with topology-aware interactions, custom coupling potentials, and stochastic noise. The resulting system of nonlinear ordinary differential equations opens a path to modeling key performance phenomena of parallel programs, including synchronization, delay propagation and decay, bottlenecks, and self-desynchronization. 

This paper introduces interaction potentials to describe memory- and compute-bound workloads and employs multiple quantitative metrics -- such as an order parameter, synchronization entropy, phase gradients, and phase differences -- to evaluate phase coherence and disruption. We also investigate the role of local noise and show that moderate noise can accelerate resynchronization in scalable applications. Our simulations align qualitatively with MPI trace data, showing the potential of physics-informed abstractions to predict performance patterns, which offers a new perspective for performance modeling and software-hardware co-design in parallel computing.
\end{abstract}




\end{frontmatter}


\section{Introduction}\label{sec:intro}

Parallel applications running on distributed-memory systems exhibit intricate performance behaviors, such as idle wave propagation and decay, process desynchronization, and spontaneous resynchronization.
An idle wave is a delay in one MPI process that ripples through the program via inter-process dependencies and fades due to noise and hardware inhomogeneities~\cite{AfzalHW19}, leading to resynchronization in resource-scalable programs \cite{AfzalEuroMPI19Poster,AfzalPoster:2021:1} but causing desynchronization and the formation of a persistent computational wavefront in resource-bottlenecked programs~\cite{AfzalHW20, AfzalHW:2022:3} if there are no frequent synchronizing collectives and the communication topology is local.
These behaviors stem from the complex interplay of computation, communication, and topology-induced mechanisms. Noise, load imbalances, hardware topology, and resource bottlenecks break the idealized translational symmetry of many parallel programs \cite{AfzalHW:2022:4}. As a result, traditional runtime and performance models -- often based on additive cost assumptions or worst-case analysis (e.g., treating overall runtime as the sum of computation and communication time) -- lack the expressiveness to capture such time-dependent effects \cite{AfzalHW:2022:1,AfzalHW:2023:1,AfzalHW:2022:4,AfzalPoster:2022:2}.

In this paper, we describe a physically inspired nonlinear model of coupled oscillators to emulate the collective dynamics of MPI processes \cite{Afzal:2023:3, AfzalISC21Poster}. The model draws its foundational analogy from the Kuramoto model \cite{kuramoto1975self, kuramoto1984chemical}, originally developed to study synchronization in natural systems. However, our formulation extends beyond Kuramoto by incorporating sparse communication topologies, custom interaction potentials, and stochastic noise, all of which are essential to mirror the heterogeneous and asynchronous nature of real-world parallel applications.

We present both a theoretical formulation of the model and a practical simulation tool designed to visualize and investigate dynamic behavior in various scenarios. The goal is to provide a lightweight and insightful way to explore, diagnose, and predict performance patterns, possibly offering a new perspective for performance modeling and software-hardware co-design in HPC.
This insight could also guide optimizations; for example, bottleneck evasion through desynchronization could enable more cost-effective hardware design choices.

\subsection{Prior contributions in \cite{Afzal:2023:3}}
In \citet{Afzal:2023:3}, the authors introduced a parameterized physical model of coupled oscillators, implemented as an interactive MATLAB tool. Inspired by the Kuramoto model, this framework enables experimentation with a system of ordinary differential equations describing the physical model.
It is particularly appealing in scenarios where communication topology and system bottlenecks significantly influence performance. The authors demonstrate that a network of coupled harmonic oscillators, with appropriately chosen interaction potentials and connectivity matrices, can reproduce some key dynamic behaviors observed in compute-bound and memory-bound parallel programs in a qualitative way. These include the propagation of idle waves, synchronization, and desynchronization phenomena.

\subsection{Contributions}
This paper builds upon and extends the work presented in~\citet{Afzal:2023:3} in several significant directions:
\begin{itemize}
    \item \emph{Better interaction potential:} We introduce refined interaction potentials to better capture phase coupling in resource-scalable and bottlenecked parallel applications, such as a Generalized Symmetric Successive Over-Relaxation (GSSOR) solver and two-dimensional five-point Jacobi method, enabling more realistic representations of synchronization and desynchronization.
    
    \item \emph{Desynchronization metrics:}
    We explore multiple observables, including the order parameter, synchronization entropy, phase gradients, and heat maps of phase differences, to provide a deeper understanding of the behavior of the system under different coupling regimes and communication delays.
    
    \item \emph{Noise as a modeling dimension:} We examine the role of stochastic noise within the oscillator model as an analog to runtime variability in MPI programs. The study shows that noise can accelerate resynchronization, demonstrating how propagating delays in MPI systems decay more rapidly in the presence of noise.
\end{itemize}

\subsection{Overview}
This paper is organized as follows:  
After the introduction, we provide a review of related work on physical modeling approaches and analogies relevant to the dynamics of MPI-parallel, bulk-synchronous and barrier-free programs in Section~\ref{sec:background}.  
Section~\ref{sec:MPImodel} introduces our MPI-inspired coupled oscillator model, including its mathematical formulation, novel interaction potentials, extensions to the classical Kuramoto model, and the MATLAB-based simulation tool used to explore system dynamics.  
In Section~\ref{sec:evaluation}, we evaluate the model through case studies and empirical comparisons, with particular emphasis on the visualizations and metrics used to analyze synchronization and desynchronization behavior.  
Finally, Section~\ref{sec:conclusion} summarizes our contributions and outlines directions for future research.

\section{Related work}\label{sec:background}

This section reviews key strands of prior work, focusing on their relevance and limitations. It begins with physical models used to understand synchronization and emergent behavior in physics, ranging from continuum systems to discrete-state abstractions. It then discusses the classical Kuramoto model, its constraints, and draws on analogies of symmetry-breaking phenomena from classical physics, such as the Mexican hat potential and Goldstone modes.

\subsection{Physical models}

Moving beyond traditional MPI simulators used for runtime prediction \cite{AfzalPoster:2024, Hoefler-loggopsim:2010, SST:2011, BigSim:2004, Dimemas:2000, xSim:2011, SimGrid:2008}, \citet{Afzal:2023:3} proposed, for the first time, a parameterized physical coupled oscillator model to represent the dynamics of MPI programs.
The use of physical models to describe synchronization and emergent behavior in parallel computing has been inspired by various domains, including statistical physics, nonlinear dynamics, and fluid mechanics. 
Instead of focusing on isolated cost components, these models help capture performance dynamics at various levels of abstraction, from microscopic interactions to macroscopic effects.

\subsubsection{Continuum models}

Continuum models, such as reaction-diffusion systems or shallow-water equations, are effective in describing bulk phenomena like wave propagation and diffusion \cite{bresch2009shallow,HUANG20067014,HUNTER1988253,drazin1989solitons,potter2021schaum}. However, their continuous formulations and reliance on partial differential equations make them ill-suited for capturing the discrete, heterogeneous, and event-driven nature of MPI-parallel applications.

\subsubsection{Discrete-state models: The Ising analogy}

The model by \citet{ising1924beitrag}, commonly used in statistical mechanics, offers a binary-state abstraction to describe local interactions and collective state transitions. It has been applied to model task dependencies and synchronization states in parallel computing \cite{Komura2012}. However, its discrete nature does not capture the continuous evolution of compute-communicate phases characteristic of HPC workloads. Our model retains the emphasis on local interactions but employs a continuous phase space to better represent rank-wise execution dynamics.

\subsubsection{Coupled oscillator models: The Kuramoto analogy}

The Kuramoto model \cite{kuramoto1975self, kuramoto1984chemical} provides a foundational framework for studying phase synchronization in networks of coupled oscillators. 
Natural oscillators that exhibit synchronization phenomena include fireflies~\cite{strogatz2003sync}, lasers~\cite{roy1994chaotic}, the Millennium Bridge~\cite{strogatz2005crowd}, metronomes~\cite{pantaleone2002synchronization}, neurons~\cite{ermentrout1991multiple}, heart cells~\cite{winfree2001geometry}, circadian rhythms~\cite{leloup1998modeling}, pacemaker cells~\cite{glass2001synchronization}, synchronized crickets~\cite{greenfield1994synchronization}, and rhythmic clapping~\cite{neda2000rhythmic}.
While analytically elegant, the original Kuramoto model assumes global sinusoidal coupling, uniform oscillator frequencies, and zero communication delay -- assumptions that diverge significantly from real HPC systems. We extend this model by incorporating sparse communication topologies, non-sinusoidal interaction potentials, heterogeneous coupling strengths, and time delays. These extensions enable our model to capture more realistic performance dynamics, including desynchronization, delay propagation, and localized bottlenecks.

\begin{figure}[t]
  \centering
  \subfloat[3D Mexican hat potential]{
    \includegraphics[width=0.45\textwidth]{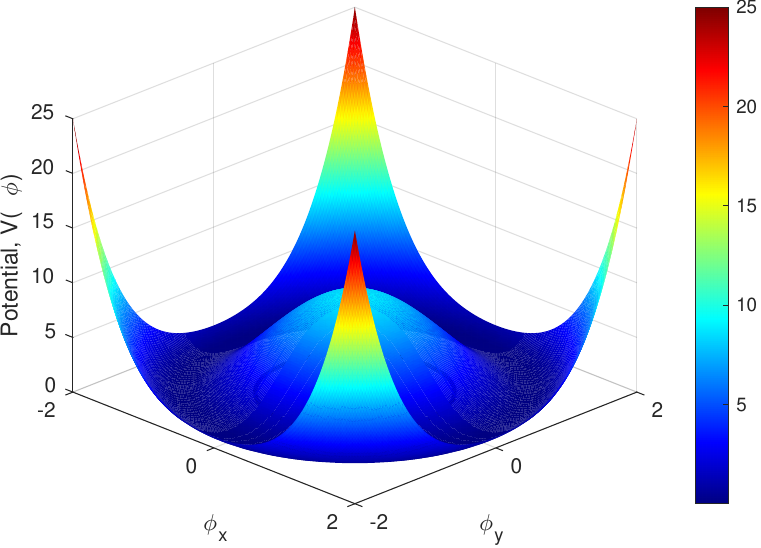}
    \label{fig:mexican_hat_3D}
  }
  \quad
  \subfloat[1D Mexican hat potential]{
    \includegraphics[width=0.45\textwidth]{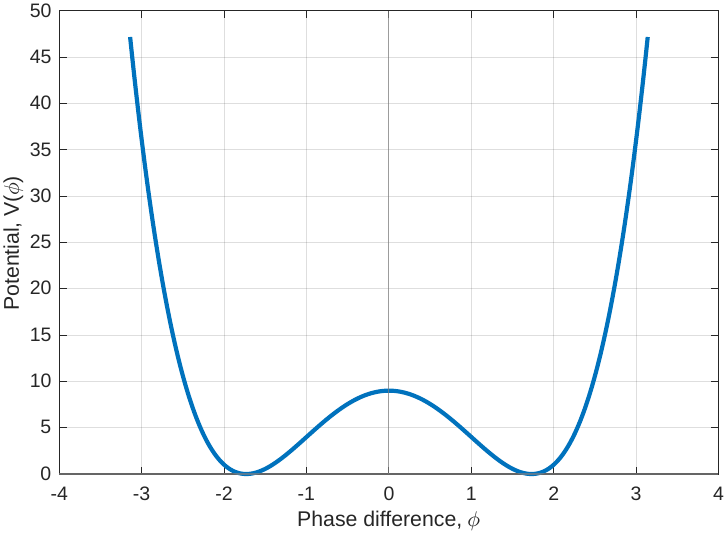}
    \label{fig:mexican_hat_1D}
  }
  \caption{Mexican hat potentials illustrating spontaneous symmetry breaking and interaction profiles in coupled oscillators. (a) 3D surface plot showing the ``ring'' of minima corresponding to broken continuous symmetry, where motion along this ring represents \emph{Goldstone modes}, low-energy excitations arising from the symmetry breaking. (b) 1D cross-section plot of potential energy as a function of phase difference, highlighting attractive and repulsive interactions.
  }
  \label{fig:mexican_hat}
\end{figure}

\subsection{Physical analogies}

Beyond explicit modeling approaches, analogies from physics offer starting points for reasoning about emergent behavior in MPI-parallel programs. These analogies allow us to frame performance patterns such as desynchronization, phase drift, and idle wave formation in terms of well-understood physical phenomena~\cite{strogatz2003sync, pikovsky2001}. By drawing parallels with force fields, energy potentials, and symmetry-breaking mechanisms, we gain intuitive and often quantitative insights into the collective dynamics of distributed systems.

\subsubsection{Newtonian interaction and potential-based dynamics}

Our model draws inspiration from Newtonian mechanics, where interactions are treated as forces derived from potential energy gradients~\cite{kuramoto1984chemical, abrams2004chimera, sethia2014chimera}. This analogy supports the use of tunable interaction potentials, such as transcendental-based potentials and Mexican hat-style profiles, which exhibit attractive and/or repulsive behavior depending on phase difference. These potentials are instrumental in modeling desynchronization mechanisms in memory-bound parallel programs.

\subsubsection{Symmetry breaking and Goldstone modes}

Emergent desynchronization mechanisms are reminiscent of spontaneous symmetry breaking in physical systems~\cite{anderson1966considerations, fradkin2013field}. In this context, minimal perturbations can lead to macroscopic orderings. The concept of Goldstone modes \cite{goldstone1961field} -- low-energy excitations associated with broken symmetries -- provides a useful lens for analyzing long-lived, coherent patterns that persist despite underlying noise.

Figure~\ref{fig:mexican_hat} illustrates the Mexican hat potential used to model spontaneous symmetry breaking in coupled oscillators \cite{cross1993pattern} \cite{aranson2002world}. 
The 3D plot in~\ref{fig:mexican_hat_3D} shows a ring of minima representing symmetry-broken states, with motion along this ring corresponding to Goldstone modes -- low-energy excitations caused by a broken continuous symmetry. 

Many data-parallel programs running on clusters show, on first sight, perfect translational invariance: Shifting MPI processes by a fixed distance within the machine (e.g., rank $i$, which was originally located on core $i$, is moved to core $\mathrm{mod}(i+k,P)$ with $P$ being the number of processes) should not change the fundamental behavior of the program because the hardware and software are also translationally invariant (functional parallelism is one of the exceptions to this). However, since we do observe desynchronization and computational wavefronts, this symmetry can be broken in the actual manifestation of the parallel run. The underlying reason is that, while the program itself may not break translational symmetry, the hardware does: For example, clusters are built from compute nodes, and these are built from multiple processing units that have internal bottlenecks such as memory bandwidth. Communication characteristics are different between different parts of the system, from intra-socket to inter-node and beyond; the network topology may also have a hierarchical structure without perfect symmetry. If a program is sensitive to these inhomogeneities (for example, by being memory bound), it may exhibit a spontaneously broken symmetry in its performance behavior.
This is the basis of the connection between physical models for symmetry breaking and parallel code.

\section{Modeling MPI programs with coupled oscillators}\label{sec:MPImodel}

In this section, we present a coupled-oscillator framework for modeling the behavior of MPI-parallel, bulk-synchronous and barrier-free programs, beginning with the classical Kuramoto model and subsequently introducing an extended, MPI-inspired formulation that accounts for realistic communication topologies, delays, and application-specific dynamics.

\subsection{The classical Kuramoto model}

The classical Kuramoto has been used to analyze phase synchronization in systems of coupled oscillators with diverse natural frequencies. It is defined as:
\begin{equation}
\dot{\theta}_i(t) = \omega_i + \frac{K}{N} \sum_{j=1}^{N} \sin(\theta_j(t) - \theta_i(t))\cma
\end{equation}
where $\omega_i$ is the intrinsic natural frequency of oscillator $i$, $\theta_i(t)$ is the phase of oscillator $i$ at time $t$, $K$ is the global coupling constant, and the coupling term involves a sinusoidal interaction with all other oscillators in the system.

This model assumes global, symmetric coupling (all-to-all communication), instantaneous influence (no time delays), and smooth sinusoidal interaction. While powerful for theoretical analysis, the Kuramoto model is limited in capturing the complexity of real-world MPI-parallel, bulk-synchronous and barrier-free programs, where communication is sparse, delays are non-negligible, and interactions are often asymmetric or state dependent.
To address these limitations, an MPI-inspired oscillator model should include a topology matrix that restricts the influence of oscillators on each other to certain pairs (just as in an MPI program not all processes commmunicate with each other), local noise to model load imbalance, time-dependent delays for the interactions, and customizable interaction potentials.  

\subsection{An MPI-inspired, Kuramoto-like oscillator model}
 
To capture the dynamic behavior of MPI-parallel programs, we keep the notion of the Kuramoto model: Each process is a harmonic oscillator with a time-dependent phase $\theta_i(t)$. This abstraction reflects the repeating compute-communicate cycles typical of distributed applications. For $P$ processes, the evolution of each process (oscillator) is governed by the following first-order differential equation:
    \bq
    \dot{\theta}_i(t) = \frac{2\pi}{t_{\text{comp}} + t_{\text{comm}}} + \zeta_i(t) + \frac{v_p}{P} \sum_{j=1}^{P} T_{ij} V_{ij}(\theta_j(t - \tau_{ij}(t)) - \theta_i(t))\eos
    \eq
This equation consists of three key components:

\paragraph{Intrinsic frequency} $\omega_i=\frac{2\pi}{t_{\text{comp}} + t_{\text{comm}}}$ represents the natural frequency of oscillator $i$, corresponding to its regular compute-communicate cycle.

\paragraph{Local noise} $\zeta_i(t)$ models local noise, capturing system noise, load imbalance, or runtime variability.
  This can be interpreted as a jitter in the oscillator's natural cycle.

\paragraph{Coupling} The final term $\frac{v_p}{P} \sum_{j=1}^{P} T_{ij} \, V_{ij} \left( \theta_j(t - \tau_{ij}(t)) - \theta_i(t) \right)$ models the influence of other oscillators on $i$ through delayed phase interactions. This term consists of four key components:
\begin{enumerate}
      \item \emph{$v_p=\frac{\beta\cdot\kappa}{t_\mathrm{comp} + t_\mathrm{comm}}$}~\cite{AfzalHW2021} is the coupling strength, which controls the magnitude of the influence exerted by oscillators on each other. It can be associated with the speed of propagating delays in parallel programs~\cite{AfzalHW2021}: Messages sent via the eager (rendezvous) protocol have $\beta = 1~(2)$. When all outstanding non-blocking MPI requests are grouped in a single \texttt{MPI\_Waitall}, $\kappa$ equals the longest communication distance; otherwise, it is the sum of all communication distances.
      \item \emph{$T_{ij}$} defines the interaction topology. If oscillator (process) $i$ is influenced by oscillator (receives messages from process) $j$, we have  $T_{ij} = 1$, and $T_{ij} = 0$ otherwise. The matrix can encode directed, undirected, sparse, or periodic communication patterns. Note that in the original Kuramoto model, $T_{ij}=1$.
      \item \emph{$V_{ij}(\cdot)$} is the interaction potential governing the influence of $j$ on $i$. Different choices, such as, trigonometric, hyperbolic, piecewise functions, yield qualitatively different synchronization and desynchronization behaviors.   
      \item \emph{$\tau_{ij}(t)$} encodes communication delays between processes $i$ and $j$. It may be constant, stochastic, or even state-dependent.
\end{enumerate}
Table~\ref{tab:model_comparison} summarizes the key conceptual and mathematical differences between the classical Kuramoto framework and our MPI-inspired model.

\begin{table}[t]
\centering
\caption{Key differences between the classical Kuramoto model and the MPI-inspired oscillator model.}
\label{tab:model_comparison}
    \begin{adjustbox}{width=0.97\textwidth}
\begin{threeparttable}
    \begin{tabular}{@{}|>{\columncolor{rowalt}}l
                    >{\columncolor{col2}}c
                    >{\columncolor{col3}}c|@{}}
    \rowcolor{rowalt}
\hline
\textbf{Feature} & \textbf{Kumamoto model} & \textbf{MPI-inspired oscillator model} \\
\hline
time delay & none & explicit delays $\tau_{ij}(t)$ \\
coupling symmetry & symmetric / all-to-all & flexible (via $T_{ij}$) \\
phase interaction & instantaneous & possibly delayed  \\
baseline frequency & natural frequency $\omega_i$ & fixed base rate from hardware time step \\
noise model & optional local noise $\zeta_i(t)$ & local noise $\zeta_i(t)$ and time-dependent delays $\tau_{ij}(t)$ \\
coupling directionality & bidirectional\tnote{*} & flexible\tnote{o} \\
interaction potential & periodic ($\sin$) & non-periodic (custom, e.g., $\tanh$, piecewise) \\
synchronization goal & global phase locking & local alignment of compute-communicate phases \\
\hline
\end{tabular}
    \begin{tablenotes}
      \footnotesize
      \item[*] \footnotesize The Kuramoto model assumes symmetric coupling -- oscillators influence each other equally.
      
      \item[o] MPI-inspired oscillator model assumes directed and topology-aware coupling -- it may be that $T_{ij} \neq T_{ji}$. For instance, oscillator $i$ receives input from oscillator $j$ if $T_{ij} \neq 0$, but oscillator $j$ does not receive input from oscillator $i$ if $T_{ji} = 0$.
    \end{tablenotes}
\end{threeparttable}
\end{adjustbox}
\end{table}

\subsection{Model potential extensions and applications}

Our formulation extends the classical Kuramoto model to study not only steady-state synchronization and desynchronization but also transient, non-equilibrium phenomena such as idle wave propagation and phase divergence. 
This is enabled by custom interaction potentials $V_{ij}(\cdot)$, which govern how the phase of one process influences another. These potentials can be tuned to capture different classes of application behavior.
Two classes of interaction potentials $V_{ij}$ have been studied so far. Figure~\ref{fig:potential-force} compares potential functions 
for scalable and bottlenecked applications, respectively. The scalable case leads to synchronization, whereas the bottlenecked case employs short-range repulsion (i.e., a negative slope around zero phase difference) to model desynchronization behavior.

\begin{table}[t]
\centering
\caption{Key differences in coupled oscillator dynamics arising from the choice of interaction potential in MPI-parallel scalable programs.}
\label{tab:potential_CB}
\begin{adjustbox}{width=0.97\textwidth}
    \begin{tabular}{@{}|>{\columncolor{rowalt}}p{3.7cm}
                    >{\columncolor{col2}}p{4.8cm}
                    >{\columncolor{col3}}p{4.8cm}
                    >{\columncolor{col4}}p{5.5cm}|@{}}
    \rowcolor{rowalt}           
\toprule
\textbf{Property} & $\boldsymbol{\sin(\theta)}$ & $\boldsymbol{\tanh(\theta)}$ & $\boldsymbol{\tanh(s\theta)}$
 \\
\midrule
{periodicity} & periodic –- cycles every $2\pi$ & non-periodic & non-periodic \\
{saturation} & none & gradual -– levels off at $\pm1$ & sharp – levels off quickly at $\pm1$ \\
{coupling smoothness} & smooth and symmetric & smooth but less symmetric & nearly binary at large $s$ \\
{rotational symmetry} & preserved -- circular phases & broken -- phase difference asymmetry & broken -- phase difference asymmetry \\
\bottomrule
\end{tabular}
\end{adjustbox}
\end{table}

\subsubsection{$tanh(s\theta)$-based potential for scalable applications} 

\begin{figure}[t]
    \centering
    \subfloat[Potential for scalable applications ($\sin$ potential shown for reference)]{
        \includegraphics[width=0.45\textwidth]{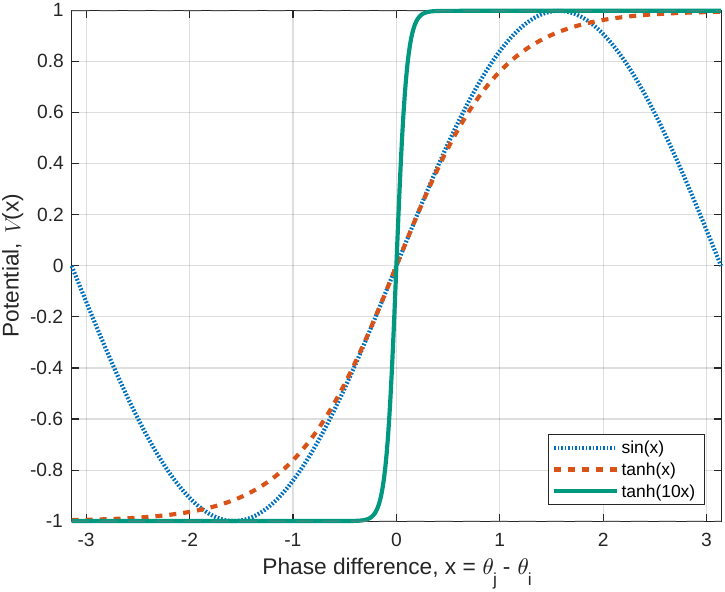}
    \label{fig:potential-CB}}
    \quad
    \subfloat[Potential for bottlenecked applications]{
        \includegraphics[width=0.45\textwidth]{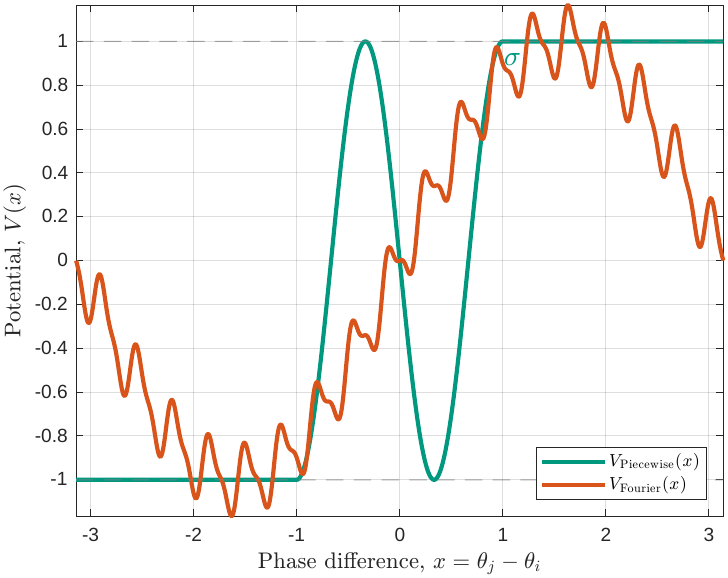}
    \label{fig:potential-MB}}
    
    \caption{Comparison of different potential functions, 
    illustrating their behavior across phase differences in the oscillator model. }
    \label{fig:potential-force}
\end{figure}

This potential is designed to model compute-bound or well-balanced parallel applications, which  tend to maintain synchronization; see Figure \ref{fig:potential-CB}.
The ${\tanh(s\theta)}$ function leads to bounded attraction due to its positive slope:
    \bq
    \label{eq:attractive}
    V (\theta_j-\theta_i) = \tanh \left( s \cdot \left(\theta_j(t) - \theta_i(t)\right)\right)\cma
    \eq
where the scaling parameter $s$ controls the steepness of the coupling response.

The interaction functions encode how strongly one process (oscillator) pulls another toward synchronization based on their phase difference. Below, we analyze the behavior of various potential functions employed in scalable parallel applications, beginning with the sinusoidal function $\sin(\theta)$ from the classical Kuramoto model, proceeding to the hyperbolic tangent $\tanh(\theta)$ as proposed in \cite{Afzal:2023:3}, and concluding with the scaled hyperbolic tangent $\tanh(s\theta)$ formulation introduced above.

\paragraph{${\sin(\theta)}$ – smooth and periodic}
The classical Kuramoto interaction promotes global phase coherence through symmetric, periodic coupling. The synchronizing influence increases smoothly with phase difference up to \(\theta = \pi\), and reverses direction beyond that. This function is well-suited for systems with homogeneous, all-to-all coupling and serves as a prototypical potential for collective synchronization. It is, however, not suitable for the description of MPI programs, since communication dependencies do not allow communicating processes to divert by more than one period. 

\paragraph{${\tanh(\theta)}$ – gentle threshold behavior} 
This function provides a saturating, non-periodic alternative to ${\sin(\theta)}$. It 
behaves like a soft threshold function, trying to sync ``far apart'' oscillators while maintaining smooth behavior near $ x = 0 $.

\paragraph{${\tanh(s\theta)}$ – hard threshold behavior} 
As $s$ increases, the function approximates a step function. Unlike ${\tanh(\theta)}$, which smoothly limits coupling across all ranges, the steep variant $({\tanh(s\theta)})$ concentrates coupling effects around a narrow window. This allows us to model synchronization dynamics in MPI programs where delays propagate locally and resynchronization is driven by tightly coupled process neighborhoods rather than global coherence.

Table~\ref{tab:potential_CB} compares the behavior of various potential function choices for scalable applications. Each function results in distinct dynamics by influencing how oscillators interact and synchronize. The sine function enables global interactions and allows for equilibrium points with a phase difference of multiples of $2\pi$, whereas $\tanh(\theta)$ leads to full synchronization with zero phase difference. The scaled $\tanh(s\theta)$ sharpens this effect further, enforcing near-neighbor coupling suited for scalable, MPI-parallel applications.

\smallskip  \highlight{\emph{Upshot 1}: In directed topologies (e.g., $T_{ij} = \delta_{i,j+1}$), oscillator $i$ is influenced by $j$ via the phase difference $\theta_j - \theta_i$, but not vice versa. The effect is unidirectional, reflecting MPI communication patterns with asymmetric dependencies.}

\smallskip  \highlight{\emph{Upshot 2}: 
Gentle $\tanh(\theta)$ slopes spread moderate coupling across a wide range, supporting global coherence. Steep $\tanh(s\theta)$ slopes focus strong corrections on large deviations, enabling fast, localized resynchronization in scalable MPI programs.
}

\subsubsection{Hybrid piecewise-sinusoidal potential for bottlenecked applications}
This potential function is tailored for modeling desynchronization in bottlenecked (e.g., memory-bound) MPI applications, where close phase alignment together with a localized communication topology maps to neighboring MPI processes being contended for a resource. To model bottleneck evasion, it introduces short-range repulsion to discourage local synchronization and long-range attraction to maintain overall phase coherence:
  \bq\label{eq:repulsive}
    V(\theta_j-\theta_i)
	= \left\{\begin{array}{lc}
      -\sin\left(\frac{3\pi}{2\sigma}\cdot(\theta_j-\theta_i)\right) & \mbox{if} \left|\theta_j-\theta_i\right|<\sigma \\
      \mbox{sgn}\left(\theta_j-\theta_i\right) & \mbox{else}
    \end{array}\right.\eos
    \eq
This piecewise formulation captures several key dynamics observed in memory-bound applications:
\begin{enumerate}
  \item For $\left|\theta_j - \theta_i\right| < \sigma$: the repulsive sine term pushes oscillators apart if they are too close in phase, evading contention.
  \item For $\left|\theta_j - \theta_i\right| \geq \sigma$: the sign function nudges distant ranks toward rough alignment, avoiding global divergence.
  \item The threshold $\sigma$ controls the width of the repulsive zone and can be tuned based on the application's sensitivity to bottlenecks.
\end{enumerate}

The potential profile (Fig.~\ref{fig:potential-MB}) exhibits a repulsive negative slope near zero and a flat attraction outside, ensuring oscillators desynchronize when too close and resynchronize when too far apart. However, since the hybrid piecewise-sinusoidal potential  introduces non-differentiability and analytical challenges due to its discontinuities, we also explored a smoother, more analytically tractable Fourier-based antisymmetric coupling function of the form:
\bq\label{eq:fourier}
V(\theta_j - \theta_i) = \sin(\theta_j - \theta_i) - a \sin(N (\theta_j - \theta_i)) + b \sin(2N (\theta_j - \theta_i))
\eq

This potential retains the desirable antisymmetry property $V(-\theta) = -V(\theta)$ and introduces higher harmonics to achieve controlled desynchronization. The fundamental term $\sin(\theta_j - \theta_i)$ promotes synchronization, but its influence can be suppressed or inverted. The $N$-th and $2N$-th harmonics induce structured desynchronization, stabilizing desynchronized states depending on the sign and magnitude of coefficients $a$ and $b$. The harmonic number $N$ is chosen to match the number of processes, ensuring that the induced phase configurations align with desirable patterns of even spacing.

This formulation is particularly well-suited to memory-bound codes, as it captures a balance between short-range repulsion and long-range synchrony using only a few harmonics. Fourier-based potential offers several advantages over the piecewise function as it is differentiable and can be analyzed via linear stability methods and spectral analysis. The coefficients $a$, $b$, and the harmonic order $N$ give precise control over the location and stability of fixed points, allowing for tailored behavior across different application topologies. Being globally defined and smooth, this potential is easier to implement in numerical solvers and avoids discontinuity-induced artifacts. However, due to phase periodicity, it allows arbitrarily large shifts between oscillators, complicating synchronization. Figure~\ref{fig:potential-MB} shows a direct visual comparison of the piecewise and Fourier-based potentials, and Table ~\ref{tab:potential-comparison} details their mathematical and computational properties.

\begin{table}[t]
\centering
\caption{A comparison of piecewise-sinusoidal and Fourier-based interaction potential functions for MPI-parallel programs experiencing bottlenecks.}
\label{tab:potential-comparison}
\begin{adjustbox}{width=0.97\textwidth}
    \begin{tabular}{@{}|>{\columncolor{rowalt}}p{4cm}
                    >{\columncolor{col2}}p{7.5cm}
                    >{\columncolor{col4}}p{7cm}|@{}}
    \rowcolor{rowalt}           
\toprule
\textbf{{Aspect}} & \textbf{{Piecewise-sin potential}} & \textbf{{Fourier-based potential}} \\
\midrule
{form} & piecewise: sine repulsion for $|x|<\sigma$, sign function otherwise & smooth sum of sines: \newline $\sin(x)-a\sin(Nx)+b\sin(2Nx)$ \\
{smoothness} & continuous, non-differentiable at $\pm\sigma$ & infinitely differentiable (smooth) \\
{local phase repulsion} &  strongly, near zero via sine segment & near zero via high harmonics  \\
{attraction behavior} & weakly, at large $|x|$ via sign   & at large $|x|$ via fundamental harmonic\\
{repulsion width} & via single threshold $\sigma$  & $a,b,n$ tune fixed points and stability \\
{numerical stability} & possible artifacts near $\pm\sigma$ & smooth, numerically stable \\
{analytical tractability} & challenging due to discontinuity & amenable to fourier-based linear stability analysis \\
\bottomrule
\end{tabular}
\end{adjustbox}
\end{table}

\subsection{Model framework}\label{sec:framework}

A lightweight, MATLAB-based solver and visualization tool \footnote{\url{https://github.com/RRZE-HPC/OSC-AD}} was designed to explore the dynamics of MPI-parallel, bulk-synchronous and barrier-free applications modeled as coupled oscillators. The framework supports multiple communication topologies -- including open chains, rings, meshes, and random graphs -- and allows adjustable boundary conditions and interaction potentials.

By tuning model parameters such as coupling strength, noise, and communication delay, the framework enables investigation of key performance phenomena including synchronization, idle wave propagation, bottleneck amplification, and symmetry-breaking effects. 

\subsubsection{Key features} \label{sec:framework:features}

The simulator provides the following configurable elements:

\paragraph{Communication topology} Users can define the adjacency matrix $T_{ij}$ to emulate diverse MPI communication patterns, including chains, rings, meshes, or random graphs.
\paragraph{Noise} Local stochastic fluctuations $\zeta_i(t)$ and communication delays $\tau_{ij}(t)$ can be tuned to simulate load imbalance, jitter, or other runtime noise.
\paragraph{Interaction potentials} Multiple coupling functions are supported, including $\tanh$-based and hybrid piecewise-sinusoidal potentials, with adjustable stiffness and range.
\paragraph{Initial condition} Simulations can be initialized from synchronized, phase-offset, or randomized states to explore both transient and long-term dynamics. The oscillator model supports multiple initial condition configurations:
    \begin{enumerate}
      \item \emph{Uniform initialization -- $\theta_i = \mathrm{zeros}(n,1)$}: All oscillators are initialized with the same phase value $\theta_i(0) = 0$, corresponding to perfect synchrony.
      \item \emph{Randomized initialization -- $\theta_i = 2\pi\cdot\mathrm{rand}(n,1)$}: Phases are drawn from a uniform distribution over $[0, 2\pi]$, simulating an initially disordered system. 
      \item \emph{Linearly spaced initialization -- $\theta_i = i/(2\pi n)$}: Phases are evenly spaced across $[0, 2\pi]$, emulating a perfect wavefront. 
      \item 
      \emph{Localized perturbation -- $\theta_i = \mathrm{zeros}(n,1);\ \theta_1\ldots\theta_k = v$}: A small subset of oscillators is perturbed; in this case, the first $k$ oscillators are initialized with phase $v$, while all remaining oscillators are initialized with zero.
    \end{enumerate}

\begin{table}[t]
    \centering
    \caption{Visualization methods and their associated measures for interpreting synchronization phenomena.}
    \label{tab:plot_comparison1}
    \begin{adjustbox}{width=0.97\textwidth}
    \begin{tabular}{@{}|>{\columncolor{rowalt}}l
                    >{\columncolor{col2}}c
                    >{\columncolor{col3}}c|@{}}
    \rowcolor{rowalt}
    \hline
    \textbf{Plots} & \textbf{Measure} & \textbf{Interpretation} \\
    \hline
    phase circle                        & \(\theta_k\) on unit circle         & visual synchronization \\
    order parameter                     & \(R(t)\) global synchrony           & synchronization level \\
    synchronization entropy             & phase distribution entropy          & disorder/order \\
    topological phase gradient                           & local phase interaction             & coupling forces \\
    pairwise phase difference timelines                & \(\Delta \theta_{ij}(t)\)           & pairwise phase locking \\
    phase difference histogram snapshots      & distribution of \(\Delta \theta_{ij}\) & snapshot of phase spread \\
    phase difference heatmap        & matrix of \(\Delta \theta_{ij}(t)\) & clusters and incoherence \\
    \hline
    \end{tabular}
    \end{adjustbox}
\end{table}

\subsubsection{Visualization options}

The solver employs a Dormand–Prince Runge-Kutta solver with adaptive time-stepping to ensure numerical stability and resolution of sharp transitions. The following visualization components are provided for analysis:
\begin{enumerate}
  \item \emph{Phase circle:} Instantaneous snapshots of oscillator phases on the unit circle to show alignment or clustering.
  \item \emph{Order Parameter:} Time-series of the Kuramoto order parameter, capturing global synchrony.
  \item \emph{Synchronization entropy:} Shannon entropy of phase distributions to detect phase scattering or multimodal clustering.
  \item \emph{Topological phase gradient:} Weighted pairwise phase differences to examine local interactions and phase tensions.
  \item \emph{Pairwise phase difference timelines:} Visualizes temporal evolution of pairwise differences, aiding in detection of phase locking or drift.
  \item \emph{Phase difference histogram snapshots:} Distribution of pairwise phase differences at selected time steps.
  \item \emph{Phase difference heatmaps:} Matrix views of pairwise differences to highlight clusters, synchronization domains, or wavefronts.
\end{enumerate}
These visualizations correspond to specific metrics -- ranging from global synchrony to local interactions -- summarized in Table~\ref{tab:plot_comparison1}, which outlines the associated measurements and their interpretations.

\section{Evaluation and case studies}\label{sec:evaluation}

To assess the effectiveness of the proposed oscillator-based model, we qualitatively compare its simulated dynamics against empirical MPI trace data collected from real application runs on HPC clusters. 
We propose and explore various metrics to evaluate their utility in revealing a multi-scale characterization of MPI process behavior -- capturing the propagation and decay of delays, as well as patterns of synchronized alignment and desynchronization -- and serving as a bridge between empirical trace data and the modeled dynamics.

\subsection{Testcases and experimental setup} \label{sec:testbed}

This section outlines the experimental setup used to evaluate the MPI-inspired oscillator model. It describes the workloads, communication topologies, noise characteristics, and initial conditions used in the evaluation. Additionally, it provides details on how empirical MPI traces were collected on the testbed for comparison.

\begin{figure}[t]
    \centering

    \subfloat[Synchronized execution of scalable applications\label{fig:lockstep}]{
        \includegraphics[width=0.46\textwidth]{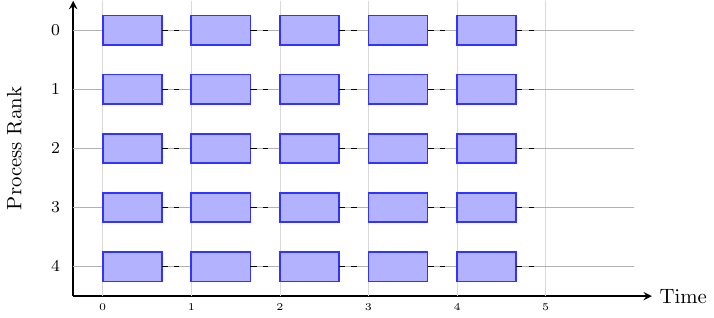}
    }
    \quad
    \subfloat[Desynchronized wavefront execution of bottlenecked applications\label{fig:wavefront}]{
        \includegraphics[width=0.46\textwidth]{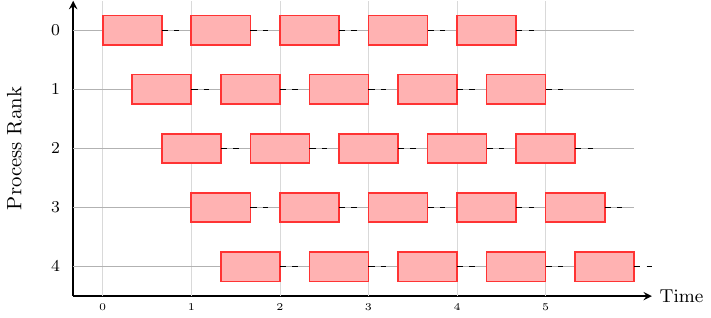}
    }

    \caption{Illustration of phase evolution across MPI ranks after absorbed perturbation. 
    (a) shows synchronized execution of scalable codes, where processes operate in lock-step after absorbing perturbations. 
    In contrast, (b) depicts a computational wavefront formed by persistent desynchronization after absorbing perturbations, often observed in memory-bound workloads. See~\cite{AfzalHW20} for details.
    }
    \label{fig:mpi-sketch}
\end{figure}

\paragraph{Workloads} We examine two representative workload classes, shown in Figure~\ref{fig:mpi-sketch}, contrasting their temporal behavior:
\begin{enumerate}
  \item \emph{Scalable (compute-bound) applications:} the GSSOR iterative solver and Pi-Solver, which maintain strong phase coherence across MPI processes when perturbations are absorbed, as shown in Fig.~\ref{fig:lockstep}.
  \item \emph{Bottlenecked (memory-bound) applications:} the 2D-5point Jacobi and STREAM Triad codes, which often exhibit phase dispersion and sustained desynchronization (computational wavefront) across MPI processes when perturbations are absorbed, as shown in Fig.~\ref{fig:wavefront}.
\end{enumerate}

\begin{figure}[t]
  \centering
  \subfloat[Bidirectional topology matrix]{
    \includegraphics[width=0.45\textwidth]{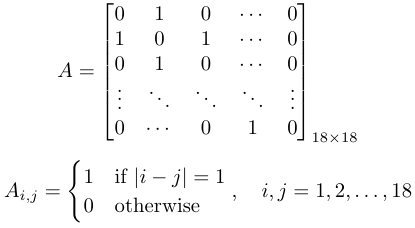}
    \label{fig:bidir_topologies}
  }
  \hfill
  \subfloat[Unidirectional topology matrix]{
    \includegraphics[width=0.45\textwidth]{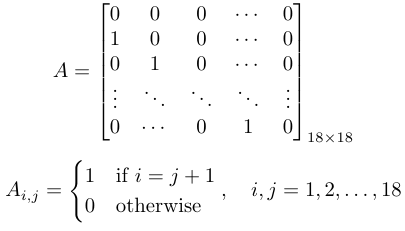}
    \label{fig:unidir_topologies}
  }
  \caption{Comparison of network topology adjacency matrices. The bidirectional topology (a) includes connections both to previous and next processes, while the unidirectional topology (b) includes connections only to the previous process, representing directed edges.}
  \label{fig:topologies}
\end{figure}

\paragraph{Communication topology} The oscillator model can be configured to accurately represent the communication structure and topology inherent to each workload. These communication patterns are encoded in the topology matrix $T_{ij}$, which may represent \emph{next-neighbor} interactions, \emph{long-range} connections, or a combination thereof. Figure~\ref{fig:topologies} presents two network topologies employing next-neighbor communication used in our experiments (Section~\ref{sec:topology}), commonly encountered in stencil solvers and iterative methods.
Subfigure~\ref{fig:bidir_topologies} depicts a bidirectional adjacency matrix with connections on both sides of the diagonal, while subfigure~\ref{fig:unidir_topologies} shows a unidirectional matrix with connections only on one side. These differences influence how information flows through the network.

\paragraph{Noise} To investigate the impact of variability on synchronization dynamics in scalable applications, we systematically varied noise parameters. For the experiment presented in Section~\ref{sec:noise}, we introduced \emph{local jitter} $\zeta_i(t)$ to model process-level variations arising from imbalance or runtime perturbations. Specifically, we applied dynamic noise terms defined as
  \bq\label{eq:noise_zeta}
    \zeta_i(t) = \frac{P_{\mathrm{noise}}}{100}\,\dot{\theta}_i(t) \cdot \mathbf{r}_i(t)\cma
  \eq
which induce mild but persistent divergence among oscillator phases over a time scale of up to 5, with $\mathbf{r}_i(t)=\mathrm{rand}(P,1)$ is a random vector. In the experiments presented in Figure~\ref{fig:unidir-sync-zeta2}, we increased ${P_{\mathrm{noise}}}$ from 100 to 2000 (expressed as a percentage of the current $\dot{\theta}_i(t))$. 

\paragraph{Initial condition} The oscillator model supports multiple initial condition configurations, as discussed in Section~\ref{sec:framework:features}. Here, we consider a special case of \emph{localized perturbation}, in which the first oscillator is assigned a phase of $\frac{3\pi}{2}$, while all others are initialized with zero phase, thereby preserving synchrony across the rest of the system (see Figures~\ref{fig:unidir-sync} to~\ref{fig:unidir-desync}).

\paragraph{Testbed and MPI trace collection}
All empirical MPI traces were collected on the \emph{Fritz} cluster at NHR@FAU\footnote{\url{https://doc.nhr.fau.de/clusters/fritz}}, which is equipped with dual-socket Intel Xeon Ice Lake processors, each comprising 36 cores, and interconnected via high-speed HDR-100 InfiniBand interconnects. 
The detailed hardware and software environment used in the experiments is summarized in Table~\ref{tab:system_Fritz}.
Sub-NUMA Clustering was enabled, dividing each socket into two ccNUMA domains of 18 cores each. All hardware prefetching mechanisms were active, while hyper-threading was disabled. MPI processes were mapped sequentially to physical cores.
The CPU frequency was fixed at its base value of 2.4\,GHz.
We consistently used the widest SIMD instruction set supported by the CPUs, which is AVX-512.
Intel Trace Analyzer and Collector (ITAC) was employed to capture performance traces. Traces were gathered by instrumenting the application code using the \texttt{-trace} option with the Intel MPI compiler wrappers. 
Upon completion, the resulting \texttt{.stf} trace files were analyzed using ITAC's graphical user interface and post-processing utilities to extract detailed process timelines.

\begin{table}[t]
    \centering
    \caption{Key hardware and software attributes of Fritz cluster.}
    \label{tab:system_Fritz}
    \begin{adjustbox}{width=0.95\linewidth}
        \renewcommand{\arraystretch}{1.1}
        \begin{tabular}{>{}c >{\columncolor{rowalt}}l l >{\columncolor{rowalt}}l l}
            \toprule
            \multirow{7}{*}{\rotatebox{90}{\textbf{~~~Architecture}}}
            & Intel Processor & Xeon Ice Lake & Processor Model & Platinum 8360Y \\
            & Base clock speed & $2.4$~\GHZ\ & Physical cores per node & 72 \\
            & ccNUMA domains per node & 4 & Sockets per node & 2 \\
            & Per-core L1/L2 cache & $48$~\KiB\ (L1) + $1.25$~\MiB\ (L2) & Shared LLC & $54$~\MiB\ (L3) \\
            & Memory per node & $4\times 64$~\GiB\ & Socket memory type & DDR4-3200 (8 ch.) \\
            & Theor. memory bandwidth & $2\times 102.4$~\GBS\ & Thermal design power & 250 W \\
            
            \midrule
            \multirow{3}{*}{\rotatebox{90}{\textbf{~~~~~~~~~Network}}}
            & Node interconnect & HDR100 Infiniband & Topology & Fat-tree \\
            & Bandwidth per link & $100$~\GBPS\ & Filesystem (capacity) & Lustre-based (3.5~PB) \\
            & I/O bandwidth & $>20$~\GBS\ & & \\

            \midrule
            \multirow{5}{*}{\rotatebox{90}{\textbf{~~~Software}}}
            & Compiler & Intel v2022u1 & SIMD & -xCORE-AVX512 \\
            & Optimization flags & -O3 -qopt-zmm-usage=high & OS & AlmaLinux v8.8 \\
            & MPI Library & Intel \verb.MPI. v2021u7 & & \\
            & ITAC version & v2021u6 & ITAC flags & -trace -tcollect \\
            \bottomrule
        \end{tabular}
    \end{adjustbox}
\end{table}

\subsection{Metrics and visualizations for synchronization and desynchronization} \label{sec:metrics}

This section presents a detailed comparative analysis of seven metrics employed to investigate synchronization and desynchronization phenomena in coupled oscillator systems. Each metric highlights unique aspects such as phase alignment, local interactions, collective behavior, and the dynamic energy landscape. Together, they enable researchers to track transitions between synchronized and desynchronized states, understand the impact of communication topology, and assess the system’s stability. Considering the variety of possible metrics \cite{AfzalHW:2022:2}, the following discussion outlines the mathematical definitions, physical interpretations, visual strengths and weaknesses, and ideal use cases for each selected metric and corresponding plot.
Throughout the section, we refer to visualizations of these metrics in Section~\ref{sec:MPIdynamicscomparison}, where comparisons between the oscillator model and MPI programs are discussed.

\subsubsection{Phase circle plot}
This plot visualizes each oscillator's phase as a point on the unit circle, offering a geometric and intuitive depiction of phase relationships. See subfigures g of Figures~\ref{fig:unidir-sync} to~\ref{fig:unidir-desync} in Section \ref{sec:MPIdynamicscomparison} for visual examples. 

\paragraph{Mathematical basis}  
Each oscillator's phase $\theta_k(t)$ is mapped to Cartesian coordinates on the unit circle:  
    \bq
    (x_k, y_k) = (\cos \theta_k(t), \sin \theta_k(t))\eos
    \eq
Of course, the time dependence of the phases cannot be captured in a static picture, but snapshots of steady-state scenarios are useful nevertheless. When viewing the phases' evolution as a video, it is advisable to use a \emph{rotating frame}, choosing an angular velocity that allows one to study the dynamics (e.g., one oscillator's current $\dot\theta_k$ or the average of all oscillators).

\paragraph{Physical interpretation}  
All points collapsed on top of each other indicate perfect \emph{synchronization}.  
Tightly grouped points at similar angles reflect partial \emph{synchronization}, while widely scattered points suggest \emph{desynchronization}.  
A uniform but ordered phase shift, such as evenly spaced points around the circle, may represent a \emph{computational wavefront}, where phases systematically increase by a fixed offset across MPI processes.

\paragraph{Strength, limitations and best suited for}
The phase circle offers an intuitive and immediate visual insight into synchrony, phase clustering, and drift. It is most effective for small to medium numbers of oscillators. For large systems, excessively overlapping points reduce clarity, making the plot crowded and difficult to interpret.

\subsubsection{Order parameter}
The order parameter $R(t)$, a widely used and easy-to-interpret scalar metric for quantifying global synchronization. Although our model generalizes beyond the classical Kuramoto formulation, the intuition provided by $R(t)$ remains useful. See subfigures b of Figures~\ref{fig:unidir-sync} to~\ref{fig:unidir-desync} in Section \ref{sec:MPIdynamicscomparison} for visual examples.

\paragraph{Mathematical basis}  
The Kuramoto order parameter is defined as:
    \bq
    R(t) e^{i\psi(t)} = \frac{1}{P} \sum_{j=1}^{P} e^{i\theta_j(t)}\cma
    \eq
where $ R(t) \in [0, 1] $ measures the degree of synchronization across all $P$ oscillators, and $\psi(t)$ is the average phase.

\paragraph{Physical interpretation} 
$R(t) \approx 1$ indicates \emph{perfect synchronization} across processes, while $R(t) \approx 0$ reflects \emph{desynchronization}. An increasing $R(t)$ signals \emph{growing synchronization}, whereas a decreasing or persistently low $R(t)$ suggests \emph{disorder}. The shape of the $R(t)$ curve can thus help differentiate between transient disruption and long-lived disorder. Beyond  $R(t)$ alone, the time derivative $\frac{dR(t)}{dt}$ can serve as an early warning metric: sharp negative slopes indicate abrupt coherence loss, while flat or positive slopes suggest recovery or stable synchronization. 
In our simulations, compute-bound applications maintained a stable $R(t)$ near unity when disturbances are absorbed, demonstrating robust synchronization. In contrast, memory-bound workloads exhibited a monotonic decline in $R(t)$, consistent with persistent desynchronization. 

\paragraph{Strength, limitations and best suited for}
The order parameter provides a concise, interpretable summary of collective behavior. While it cannot capture multi-cluster structures or partial synchronization, it remains a reliable tool for tracking the onset and degree of synchronization.

\subsubsection{Synchronization entropy}
The Shannon entropy of the oscillator phase distribution offers a complementary perspective to the order parameter by quantifying the degree of disorder and detecting multimodal or clustered phase distributions. Table~\ref{tab:sync-metrics} contrasts $R(t)$, a measure of global phase locking, with $S(t)$, which detects structural complexity such as phase clustering and multimodality. See subfigures c of Figures~\ref{fig:unidir-sync} to~\ref{fig:unidir-desync} in Section \ref{sec:MPIdynamicscomparison} for visual examples.

\begin{table}[t]
\centering
\caption{Comparison between order parameter and synchronization entropy.}
\label{tab:sync-metrics}
\begin{adjustbox}{width=0.97\textwidth}
    \begin{tabular}{@{}|>{\columncolor{rowalt}}p{4.7cm}
                      >{\columncolor{col2}\centering\arraybackslash}p{6.5cm}
                      >{\columncolor{col3}\centering\arraybackslash}p{6.1cm}|@{}}
    \rowcolor{rowalt}
    \toprule
    \textbf{Aspect} & \textbf{Order parameter } & \textbf{Synchronization entropy $\boldsymbol{S(t)}$} \\
        \midrule
    mathematical formulation & 
    \(\displaystyle R(t) e^{i\psi(t)} = \frac{1}{P} \sum_{j=1}^{P} e^{i\theta_j(t)}\) & 
    \(\displaystyle S(t) = -\sum_{k=1}^{N_b} p_k(t) \log p_k(t)\) \\[1ex]
    
    value range & 
    $R(t) \in [0, 1]$ & 
    $S(t) \in [0, \log N_b]$ (with $N_b$ bins) \\
    
    multimodality detection & 
    no & 
    yes \\
    
    measures & 
    phase alignment & 
    phase distribution spread \\
    
    interpretation & 
    $R \approx 1$: strong synchronization \newline  $R \approx 0$: desynchronization \newline $R \approx 0$: multimodal structure & 
    $S \approx 0$: strong synchronization \newline $S \approx \log N_b$: desynchronization \newline $S < \log N_b$: multimodal structure \\  
    
    sensitivity & 
    mean vector direction (1st moment) & 
    distribution shape (all moments) \\
        
    \bottomrule
    \end{tabular}
\end{adjustbox}
\end{table}

\paragraph{Mathematical basis}  
The Shannon entropy is a sum over binned oscillator phases:
    \bq
    S(t) = -\sum_{k=1}^{N_b} p_k(t) \log p_k(t)\cma
    \eq
where $ S(t) \in [0, \log N_b] $. Here, $p_k(t)$ is the normalized count of oscillators in bin $k$ at time $t$, and $N_b$ is the number of bins. 
To compute the bins adaptively, we use the Freedman–Diaconis rule: The bin width is given by $h = 2 \times \frac{\mathrm{IQR}}{N^{1/3}}$, where IQR is the interquartile range and $N$ is the number of data points. 
In practice, a small $\epsilon$ is added to the argument of the $\log$ to avoid evaluating $\log(0)$.

\paragraph{Physical interpretation}  
High entropy indicates \emph{desynchronization} (disorder), while low entropy reflects tightly clustered phases (\emph{synchronization} ). 
Entropy reflects the spread of phase distributions: $ S(t) \approx 0 $ suggests nearly all phases are aligned (strong synchronization), $ S(t) \approx \log N_b $ implies phases are uniformly spread (complete desynchronization), and Intermediate values indicates partial or clustered synchronization.
In practice, entropy can reveal structure not captured by the order parameter. For instance, in simulations with hybrid or asymmetric topologies, $R(t)$ may remain stable while $S(t)$ increases -- signaling divergence into multiple synchronized subgroups. This makes entropy especially useful for identifying multimodal or partial desynchronization states. 
Quantized or burst-like entropy values may arise from abrupt transitions between synchronized (low-entropy) and desynchronized (high-entropy) states. Each distinct entropy level often corresponds to the emergence of phase clusters. Moreover, the discrete nature of the system -- particularly due to the finite number of oscillators -- constrains the entropy to a limited set of possible values.


\paragraph{Strength, limitations and best suited for}
Entropy offers a fine-grained view of synchronization beyond what $R(t)$ captures, as it is sensitive to the internal structure of the phase distribution. Although it may be less intuitive than $R(t)$ and depends on binning choices, it is good for detecting multimodal distributions and transitions.

\subsubsection{Topological phase gradient}
This metric is a measure of how far each oscillator's phase is, on average,  away from all others it interacts with. See subfigures d of Figures~\ref{fig:unidir-sync} to~\ref{fig:unidir-desync} in Section \ref{sec:MPIdynamicscomparison} for visual examples.

\paragraph{Mathematical basis}  
For oscillator $i$, the metric is computed as the weighted sum of phase differences:  
    \bq
    g_i(t) = \sum_{j} T_{ij} \left|\theta_j(t) - \theta_i(t)\right|\cma
    \eq
where $T_{ij}$ is the communication topology matrix and $|\theta_j(t) - \theta_i(t)|$ quantifies the phase misalignment between oscillator $i$ and its neighbor $j$. This expression can be extended to include delayed interactions when applicable.

\paragraph{Physical interpretation}
The local gradient $g_i(t)$ reflects the net phase spread between oscillator $i$ and its neighbors. 
High values indicate strong phase disparities with neighbors. Low or constant values suggest steady coupling influences. 
Large fluctuations often correspond to transient dynamics or phase slips.
While a synchronized state yields a spread close to zero, a desynchronized system causes the average spread of $g_i(t)$ across oscillators tends toward a constant non-zero value over time, indicating persistent phase differences and stable desynchronization.

\paragraph{Strength, limitations and best suited for}
This metric offers a scalar diagnostic of desynchronization and can reveal asymmetries in communication, boundary effects, local dynamics or coupling imbalances across individual oscillators, which are not captured by global metrics like $R(t)$ or $S(t)$. 
When plotted as a timeline, this metric can be useful for analyzing fine-grained interactions or topology influence in small to medium-sized networks. However, visual clutter may limit interpretability in larger systems. 

\subsubsection{Pairwise phase difference timelines}

The time evolution of all pairwise phase differences offers a detailed view of synchronization and desynchronization patterns between oscillator pairs. See subfigures e of Figures~\ref{fig:unidir-sync} to~\ref{fig:unidir-desync} in Section \ref{sec:MPIdynamicscomparison} for visual examples.

\paragraph{Mathematical basis}
Each line in the plot corresponds to the phase difference between a pair of oscillators over time:
    \bq
    \{\Delta \theta_{ij}(t) = \theta_j(t) - \theta_i(t) \mid j \le i\} \eos
    \eq

\paragraph{Physical interpretation} 
In a timeline plot of $\Delta\theta_{ij}$, stable horizontal lines indicate that the oscillators $i$ and $j$ are phase-locked. Time-varying curves reflect desynchronization, fluctuations, slips, or gradual phase drift between pairs. Smooth trajectories typically correspond to stable interaction regimes, while jagged or spread-out lines imply desynchronization or chaotic dynamics.

\paragraph{Strength, limitations and best suited for}
Pairwise difference timelines offer detailed insight into the local dynamics between oscillator pairs, capturing phenomena such as phase slips, drift, or subgroup divergence that global metrics like $R(t)$ or $S(t)$ may overlook. While highly informative, their $\mathcal{O}(N^2)$ computational and visual complexity makes them less practical for large networks, where plots become dense and hard to interpret. They are best suited for small to moderate-sized systems or for targeted analysis of selected, critical pairs in larger networks.

\subsubsection{Pairwise phase difference histogram}
Pairwise phase differences between oscillators can be visualized as a histogram at a fixed point in time. See subfigures h of Figures~\ref{fig:unidir-sync} to~\ref{fig:unidir-desync} in Section \ref{sec:MPIdynamicscomparison} for visual examples.

\paragraph{Mathematical basis}
The histogram is constructed from the set of pairwise phase differences at time $t$:
\begin{equation}
\{\Delta \theta_{ij}(t) = \theta_j(t) - \theta_i(t) \mid j \le i\}
\end{equation}

\paragraph{Physical interpretation}  
The shape of the histogram reflects the degree of synchronization: A sharp, narrow peak near zero indicates strong global synchronization, whereas a tall but broader peak suggests partial synchronization or clustering. A flat or wide distribution corresponds to desynchronization.

\paragraph{Strength, limitations and best suited for}
Pairwise phase difference histograms provide a simple and interpretable distribution overview for comparing the degree of synchronization across different time points.
However, they lack temporal resolution and do not reveal dynamic behaviors such as phase slips or drift. These plots are best suited for snapshot-based analyses or for identifying changes in overall synchrony over time when used in conjunction with time-series data.

\subsubsection{Pairwise phase difference heatmap}
Pairwise phase differences can be visualized directly as a matrix heatmap at a specific point in time, visually highlighting clusters or block structures. It is effective for detecting structure like wavefronts; however, it provides only a static representation and does not capture temporal dynamics unless the time evolution is rendered into a video. See subfigures i of Figures~\ref{fig:unidir-sync} to~\ref{fig:unidir-desync} in Section \ref{sec:MPIdynamicscomparison} for visual examples.

\paragraph{Mathematical basis}  
The heatmap is constructed from the matrix of wrapped pairwise phase differences at time $t$:
\bq
M_{ij}(t) = \theta_j(t) - \theta_i(t)
\eq

\paragraph{Physical interpretation}  
The visual structure of the heatmap reflects phase relationships:
symmetric and uniform color blocks indicate \emph{synchronization}, while distinct block patterns of random color imply \emph{desynchronization}.
If all oscillators have uniform phase differences, forming a linear increase in phase across indices, essentially,
\[
\theta_j(t) - \theta_i(t) \propto j - i,
\]
then the heatmap of pairwise phase differences will exhibit a gradient-like pattern along the diagonals. 
Specifically, the first upper and lower diagonals correspond to next-neighbor phase differences, the second to next-to-next neighbors, etc. In case of a wavefront pattern, as the distance $|i - j|$ increases, the phase difference grows proportionally. Optionally, the value can be wrapped to the interval $[-\pi,pi)$ (MATLAB $\mathrm{wrapToPi}$ function).

\paragraph{Strengths, limitations, and best suited for}
Pairwise phase difference heatmaps offer a compact and intuitive visualization, making them well suited for detecting phase clusters, structural phase relationships, or wavefront-like patterns at specific time points even for large systems. However, they are static, and visual clarity may degrade, especially when individual relationships are densely encoded, unless the data is reordered or interactive zooming is available.

\begin{table}[t]
\centering
\caption{Summary of visualization plots for MPI-inspired oscillator model with their relative strengths, limitations, and recommended applications.}
\label{tab:plot_comparison2}
\begin{adjustbox}{width=0.97\textwidth}
\begin{tabular}{@{}|>{\columncolor{rowalt}}l
                >{\columncolor{col1}}c
                >{\columncolor{col2}}c
                >{\columncolor{col3}}c
                >{\columncolor{col4}}c|@{}}
\rowcolor{rowalt}
\toprule
\textbf{Plots} & \textbf{Visual clarity} & \textbf{Quantitative} & \textbf{Scalability} & \textbf{Best use case} \\
\midrule
phase circle                        & very strong         & no                    & medium        & small systems, phase patterns \\
order parameter                     & very strong         & yes                   & large         & global synchronization measure \\
synchronization entropy             & medium (less intuitive) & yes               & large         & detecting phase disorder and clustering \\
topological phase gradient         & medium              & yes (local)           & medium        & local interaction and topology analysis \\
phase differences timelines         & strong detail       & no                    & very low      & detailed pairwise locking/slips \\
phase differences histogram         & medium              & yes (snapshot)        & medium        & statistical view of synchrony at a moment \\
phase differences heatmap           & strong for clustering & no (qualitative)    & medium        & visualizing phase clusters or wavefronts \\
kuramoto potential                  & medium              & yes                   & medium        & energy landscape, stability monitoring \\
\bottomrule
\end{tabular}
\end{adjustbox}
\end{table}

\subsubsection{Potential ``energy''}

\begin{figure}[hp]
    \centering
    \subfloat[MPI trace (\href{https://github.com/RRZE-HPC/OSC-AD/blob/main/MPI_trace_GSSOR/GSSOR_d-1.mp4}{video})]{\includegraphics[scale=0.8]{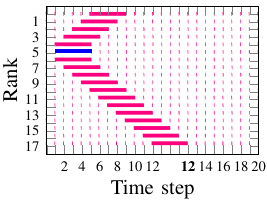}}\quad
    \subfloat[Order parameter]{\includegraphics[scale=0.25]{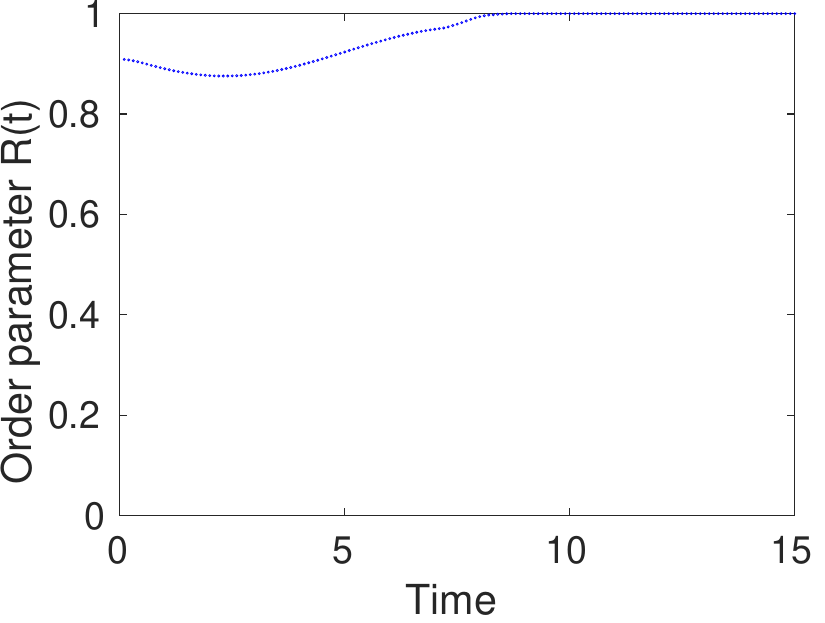}}\quad
    \subfloat[Entropy]{\includegraphics[scale=0.25]{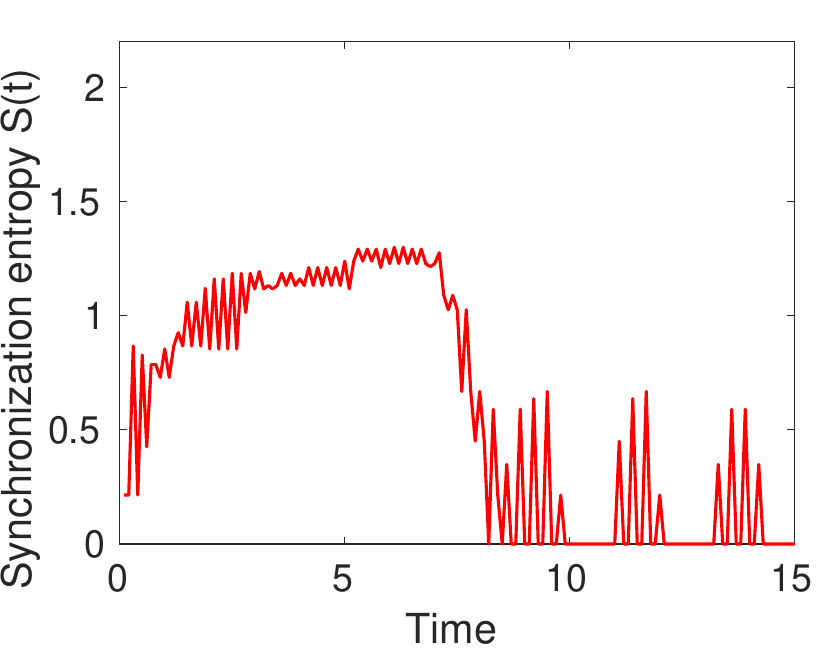}}\\[1ex]
    
    \subfloat[Phase gradient]{\includegraphics[scale=0.25]{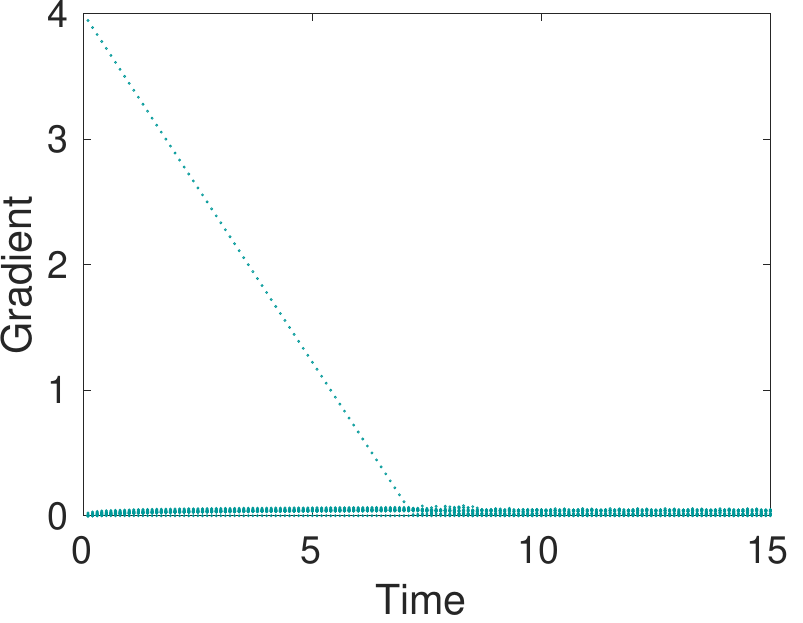}}\quad
    \subfloat[Phase differences]{\includegraphics[scale=0.25]{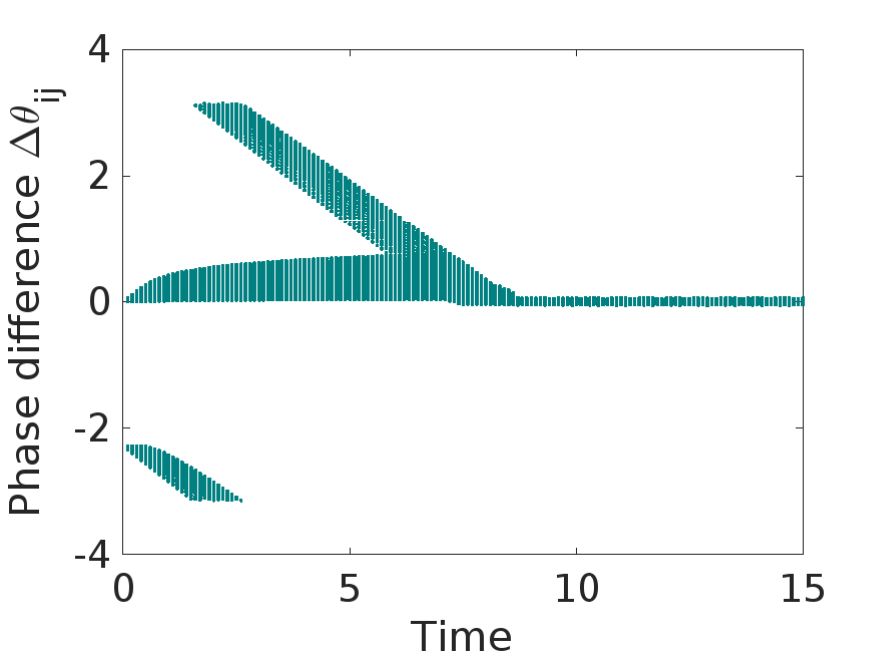}}\quad
    \subfloat[Potential]{\includegraphics[scale=0.25]{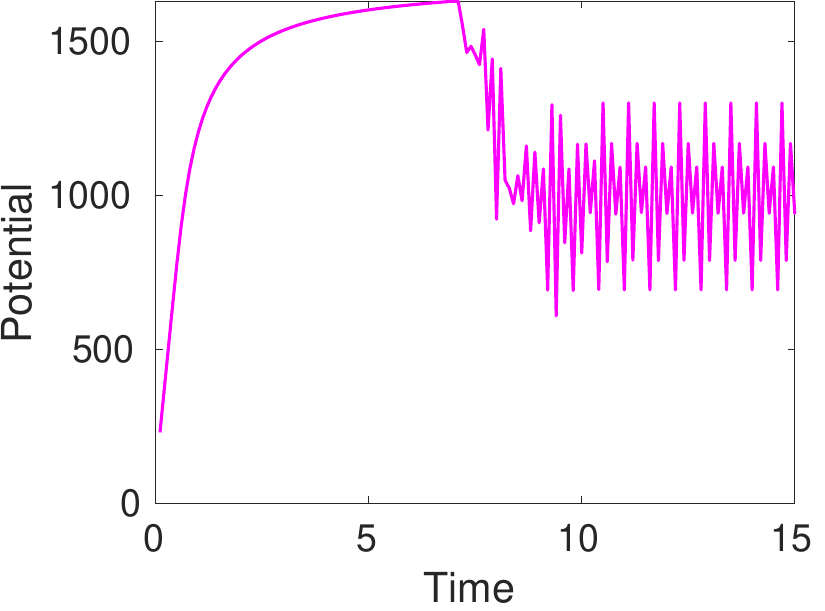}}\\[1ex]

     \subfloat[Phase circle]{
    \begin{minipage}[c]{\textwidth}
            \centering
            \includegraphics[scale=0.19]{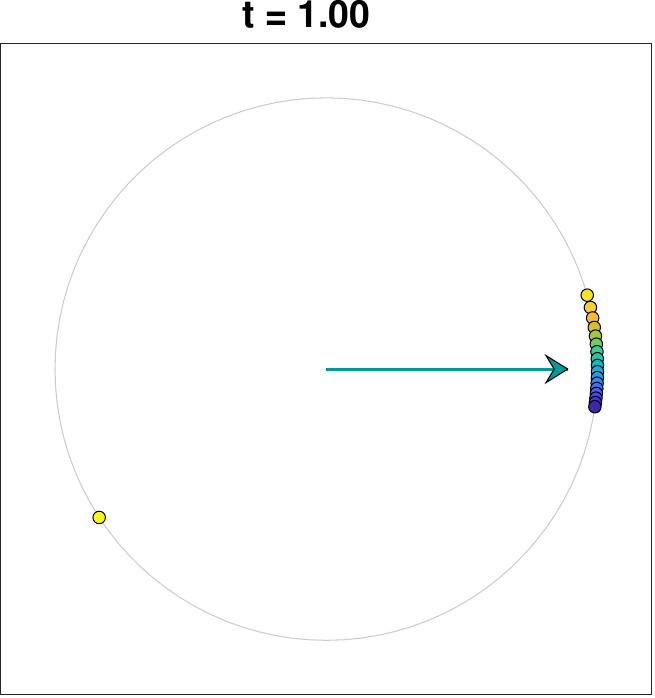}\qquad
            \includegraphics[scale=0.19]{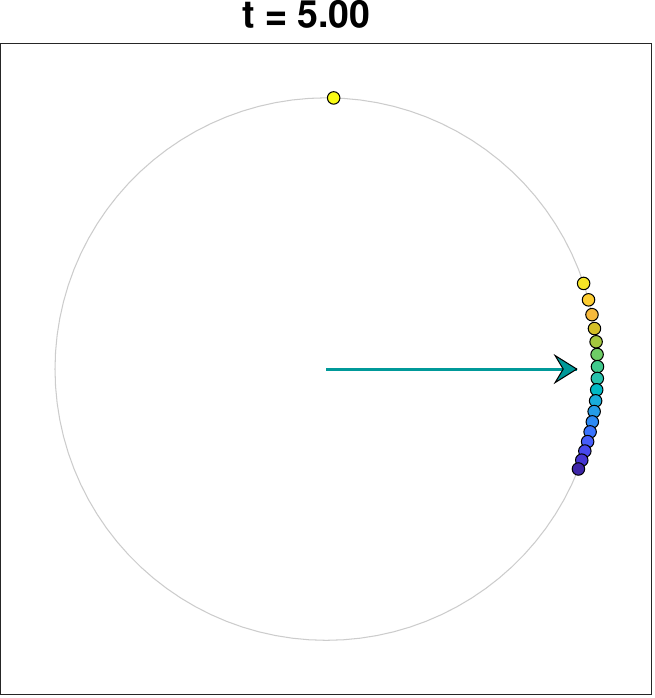}\qquad
            \includegraphics[scale=0.19]{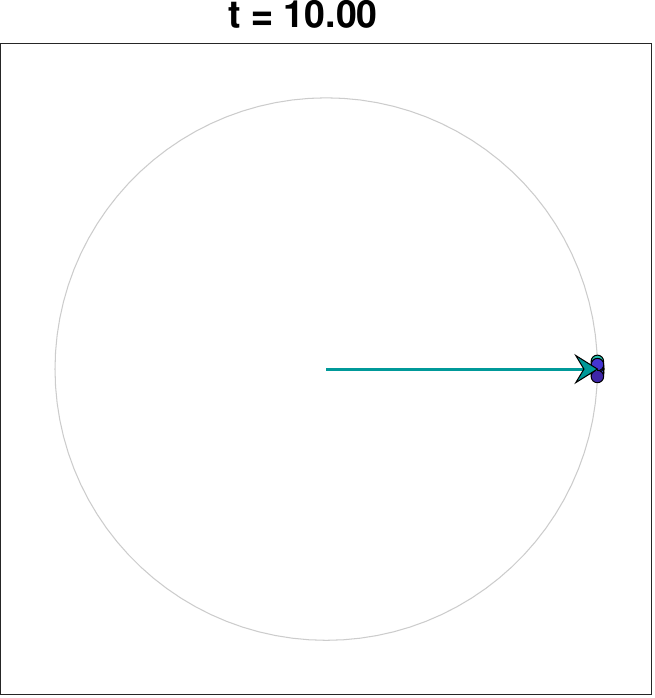}\qquad
            \includegraphics[scale=0.19]{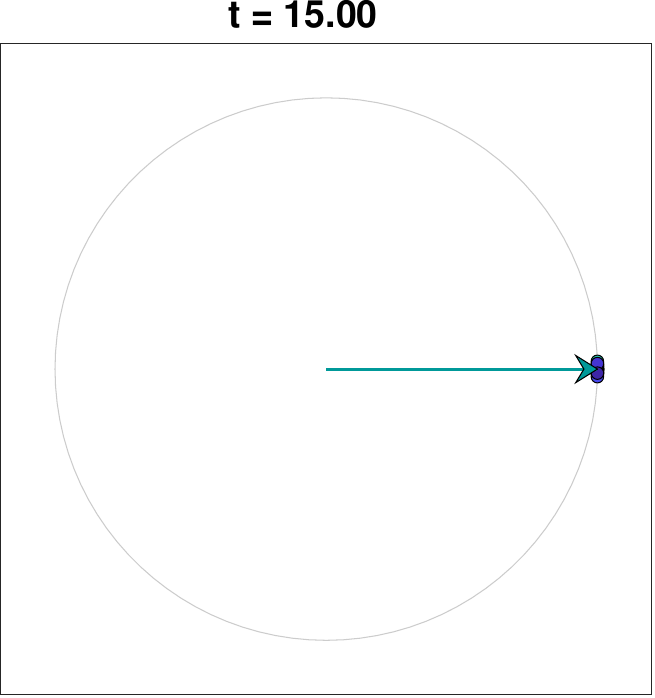}
        \end{minipage}
    }
    
     \subfloat[Pairwise phase differences histogram]{
    \begin{minipage}[c]{\textwidth}
            \centering
            \includegraphics[scale=0.2]{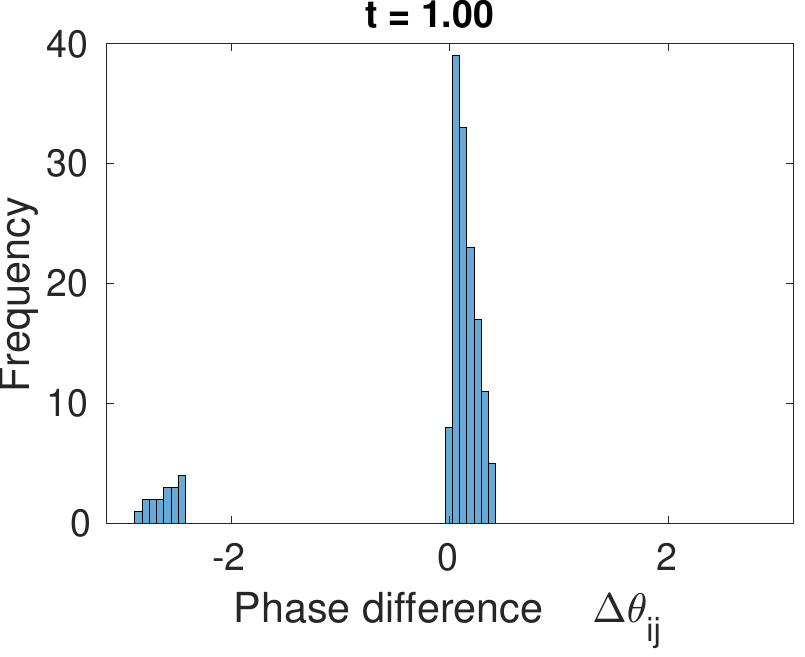}\quad
            \includegraphics[scale=0.2]{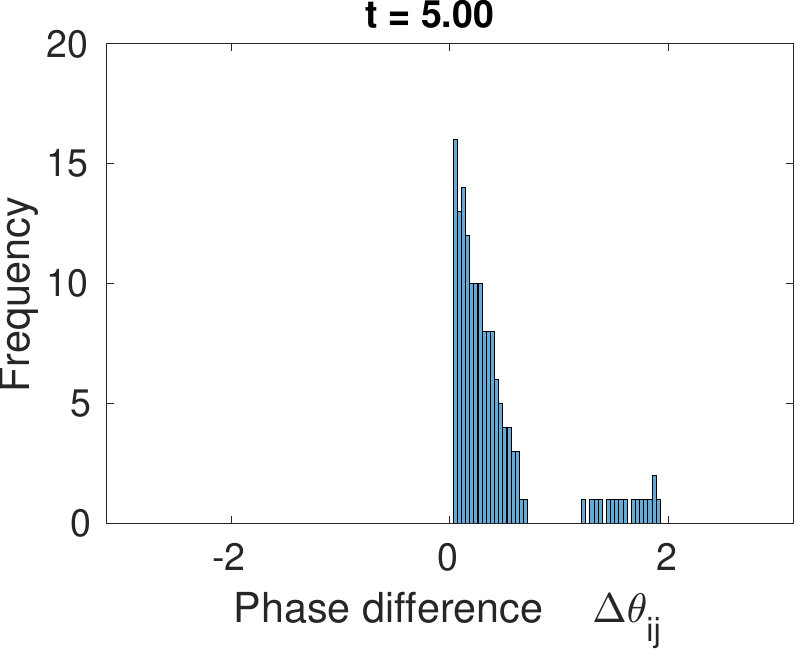}\quad
            \includegraphics[scale=0.2]{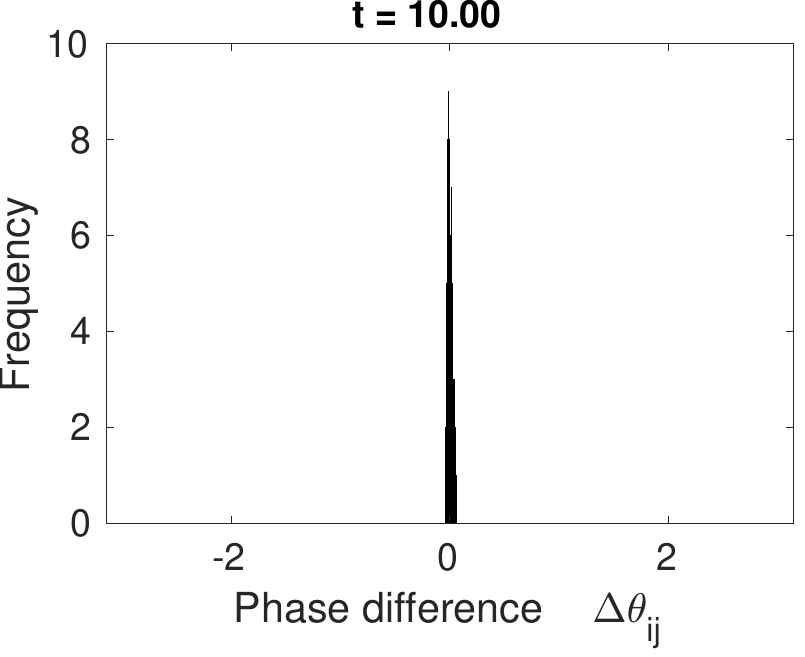}\quad
            \includegraphics[scale=0.2]{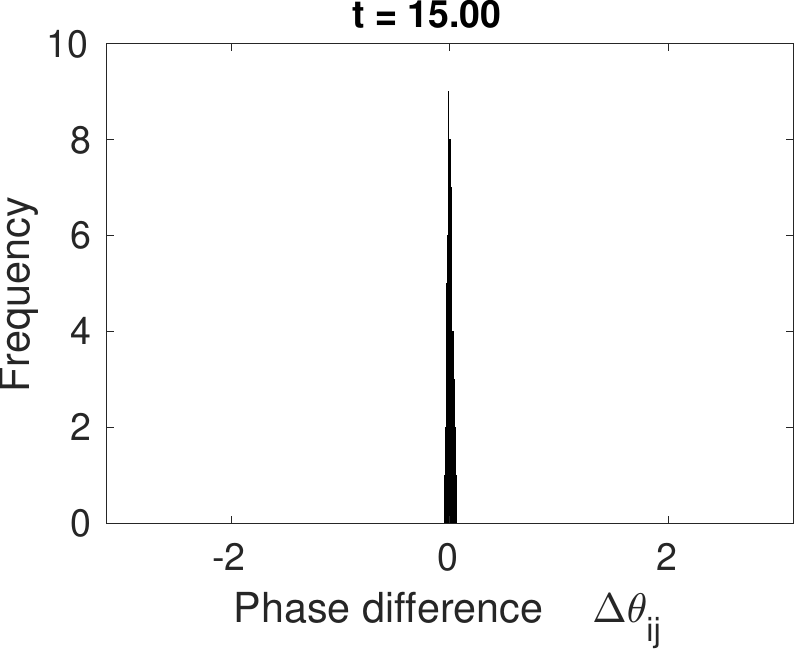}
        \end{minipage}
    }
    
    \subfloat[Pairwise phase differences heatmap]{
    \begin{minipage}[c]{\textwidth}
            \centering
            \includegraphics[scale=0.21]{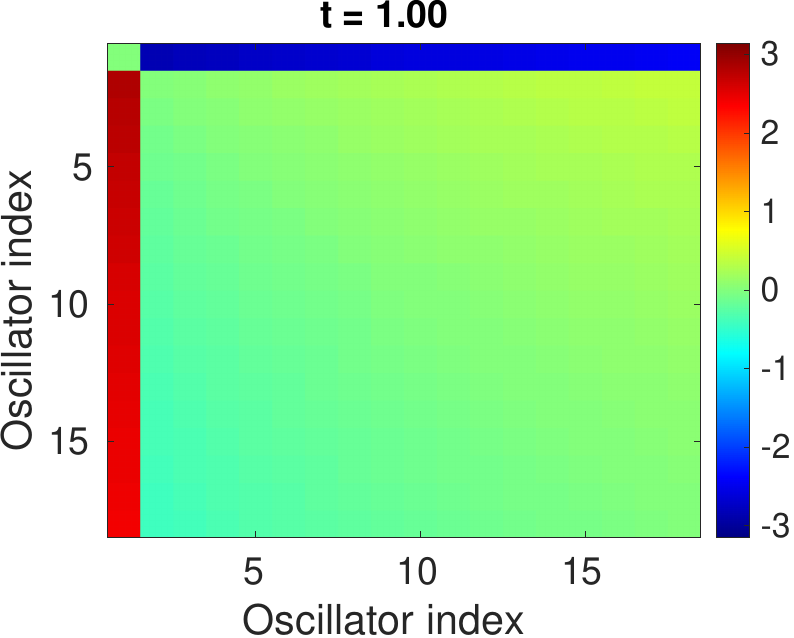}\quad
            \includegraphics[scale=0.21]{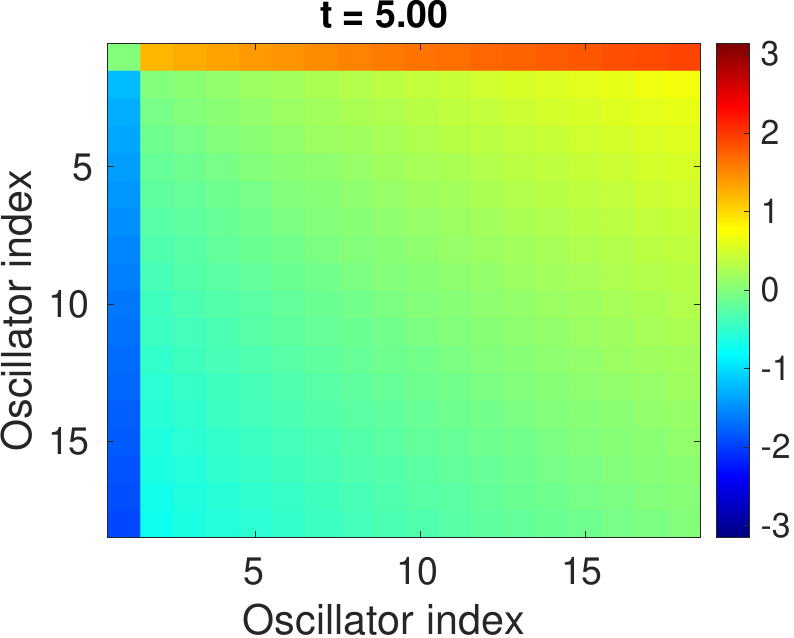}\quad
            \includegraphics[scale=0.21]{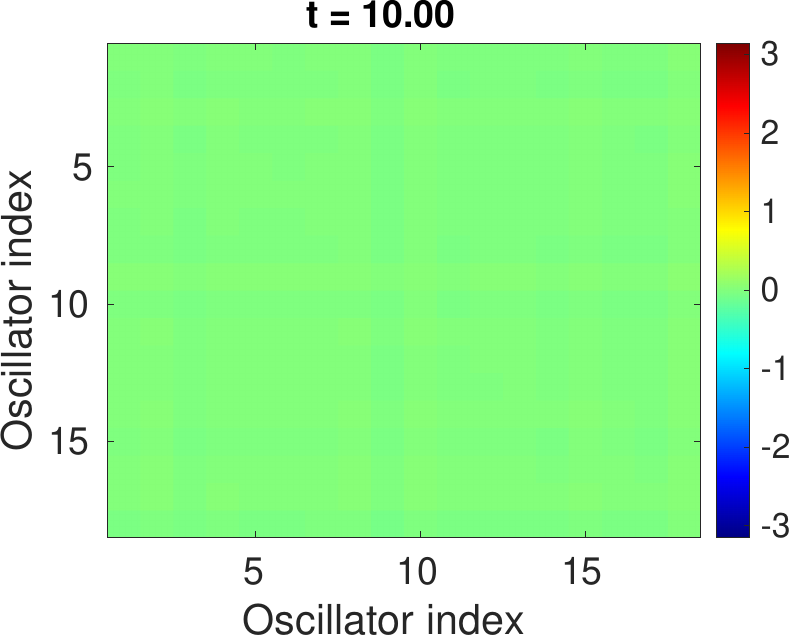}\quad
            \includegraphics[scale=0.21]{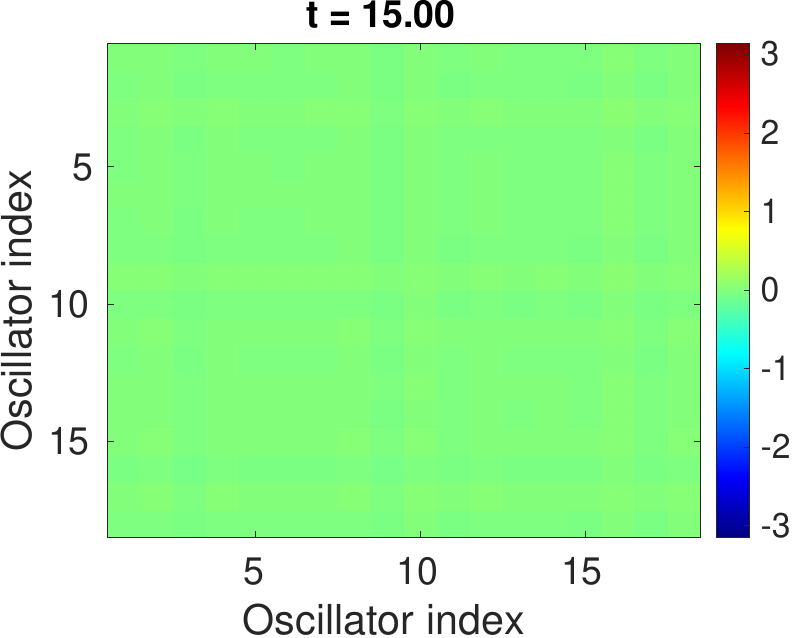}
        \end{minipage}
    }

    \caption{
        Visualization plots of the oscillator model show resynchronization analysis across 18 processes (oscillators) in GSSOR code after inducing a delay on a single process. Communication follows a \emph{unidirectional, non-periodic} topology, as illustrated in Figure~\ref{fig:unidir_topologies} and uses rendezvous protocol.
    }
    \label{fig:unidir-sync}
\end{figure}

The time-dependent potential ``energy'' quantifies the system's average ``restoring force,'' providing insight into synchronization, desynchronization, and stability. 
Note that we use the terms ``energy'' and ``force'' in a hand-waving manner here, since the oscillator model does not follow Newtonian mechanics. See subfigures f of Figures~\ref{fig:unidir-sync} to~\ref{fig:unidir-desync} in Section \ref{sec:MPIdynamicscomparison} for visual examples.

\paragraph{Mathematical basis}
Our definition of potential energy is a generalization of the \emph{Kuramoto potential}: 
\begin{equation}
V(t) = \sum_{i=1}^{P} \sum_{j=1}^{P} T_{ij} \cdot V^2 \left( \theta_j(t) - \theta_i(t) \right)\eos
\end{equation}

\paragraph{Physical interpretation}
High values of $V(t)$ indicate transient states, since nonzero values of the potential will lead to changing oscillator frequencies. A decreasing trend suggests convergence towards a stable state, whereas sudden spikes may signal instabilities or phase slips. Smooth decay to a minimum corresponds to stable phase locking, while oscillations may indicate complex dynamical states. Tracking $V(t)$ over time is particularly useful for monitoring system stability, convergence towards synchrony, dynamic fluctuations, and identifying phase slips or metastable states.

\paragraph{Strengths, limitations, and best suited for}
The potential is a theoretically grounded measure that provides valuable insight into the system's energy landscape and stability even for very large systems. However, it is more abstract than direct phase or order parameters. It is best suited for identifying energy minima and conducting time-resolved studies of synchronization stability.

\smallskip \highlight{\emph{Upshot}: 
The choice of metric and visualization -- whether combined or selectively applied -- depends on the research objective and system scale. For global synchronization metrics, the order parameter is most effective. For detailed structural insights, pairwise phase difference (timelines and heatmap) are informative but scale poorly. The potential energy offers theoretical insights into system stability, while Shannon entropy provides complementary disorder information. Phase circle plots give an immediate visual grasp of phase coherence, especially for smaller systems. For a consolidated overview of the strengths, limitations, and ideal use cases of each method, see Table~\ref{tab:plot_comparison2}. 
}

\subsection{MPI trace and oscillator model comparison}\label{sec:MPIdynamicscomparison}
\begin{figure*}[hp]
    \centering
    \subfloat[MPI trace (\href{https://github.com/RRZE-HPC/OSC-AD/blob/main/MPI_trace_GSSOR/GSSOR_d+-1.mp4}{video})]{\includegraphics[scale=0.85]{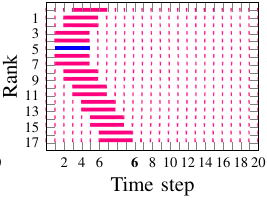}}
    \subfloat[Order parameter]{\includegraphics[scale=0.27]{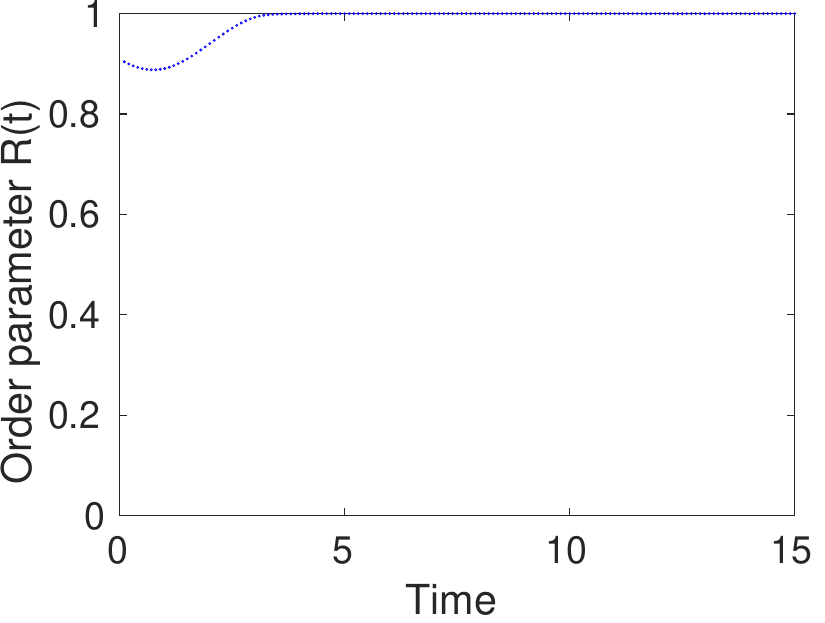}}\quad
    \subfloat[Entropy]{\includegraphics[scale=0.27]{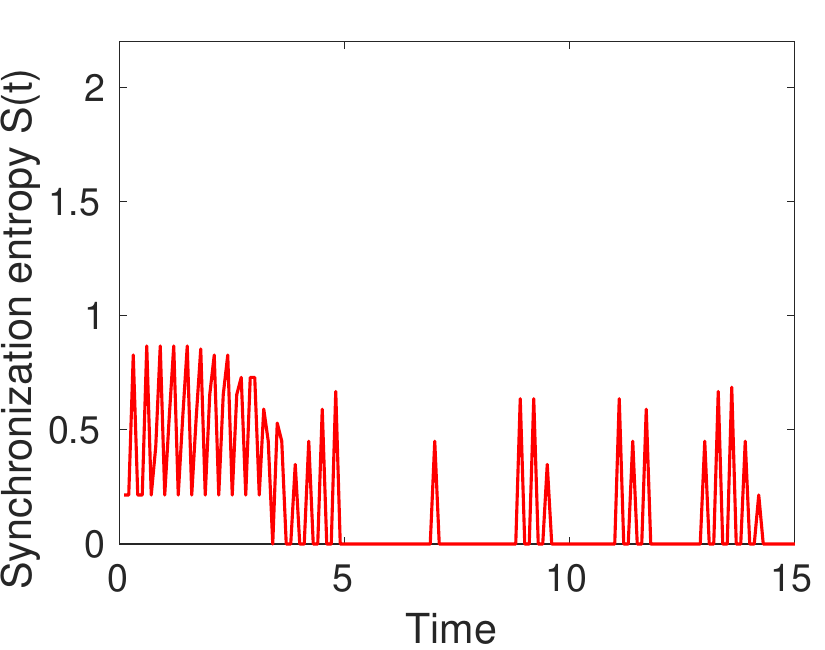}}\\[1ex]
    
    \subfloat[Phase gradient]{\includegraphics[scale=0.26]{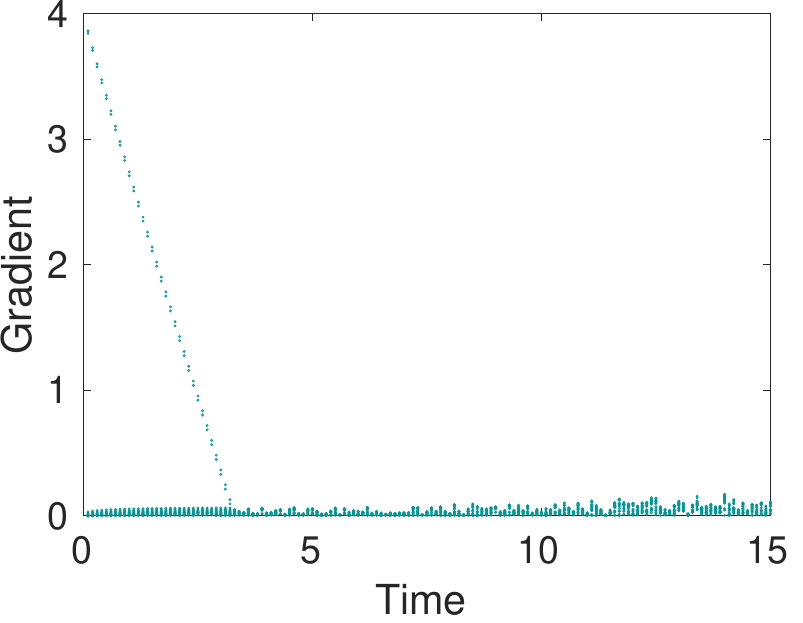}}\quad
    \subfloat[Phase differences]{\includegraphics[scale=0.26]{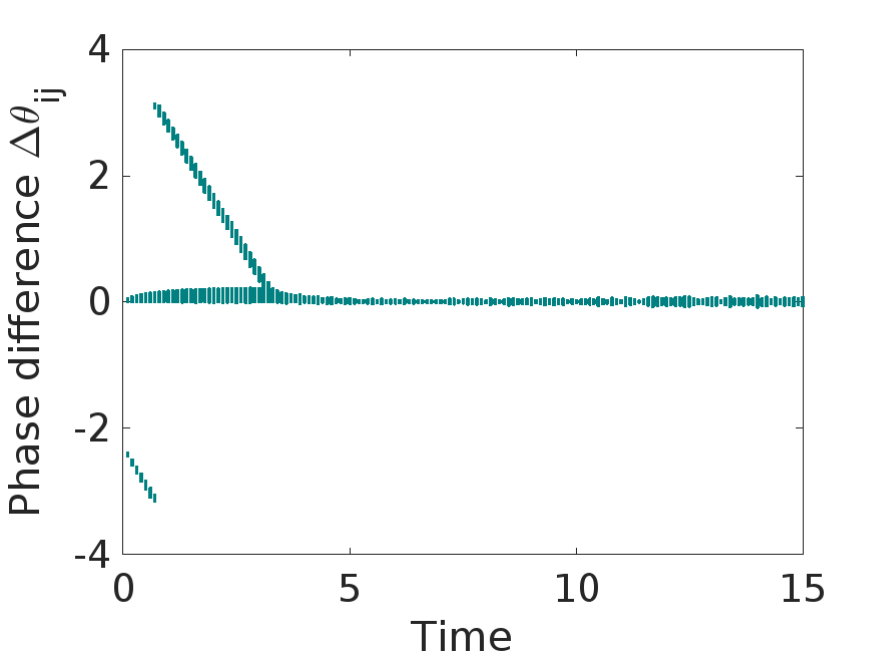}}\quad
    \subfloat[Potential]{\includegraphics[scale=0.26]{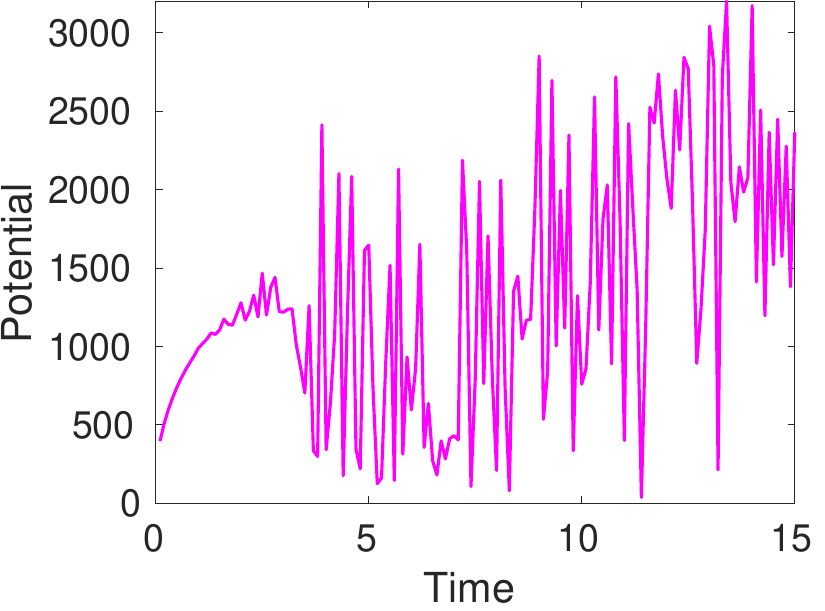}}\\[1ex]

     \subfloat[Phase circle]{
    \begin{minipage}[c]{\textwidth}
            \centering
            \includegraphics[scale=0.2]{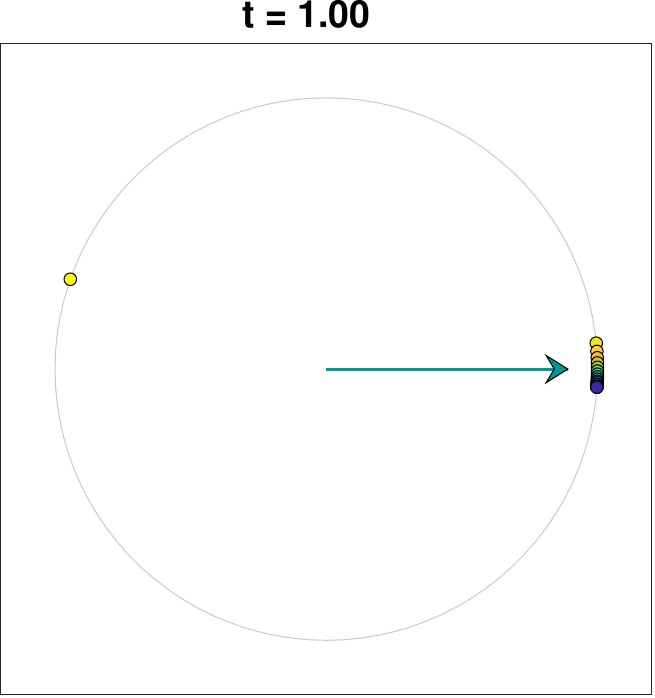}\qquad
            \includegraphics[scale=0.2]{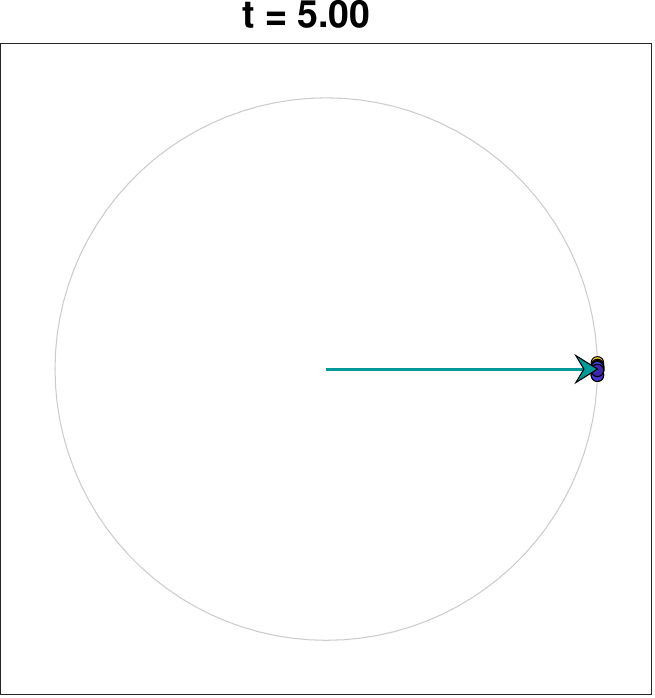}\qquad
            \includegraphics[scale=0.2]{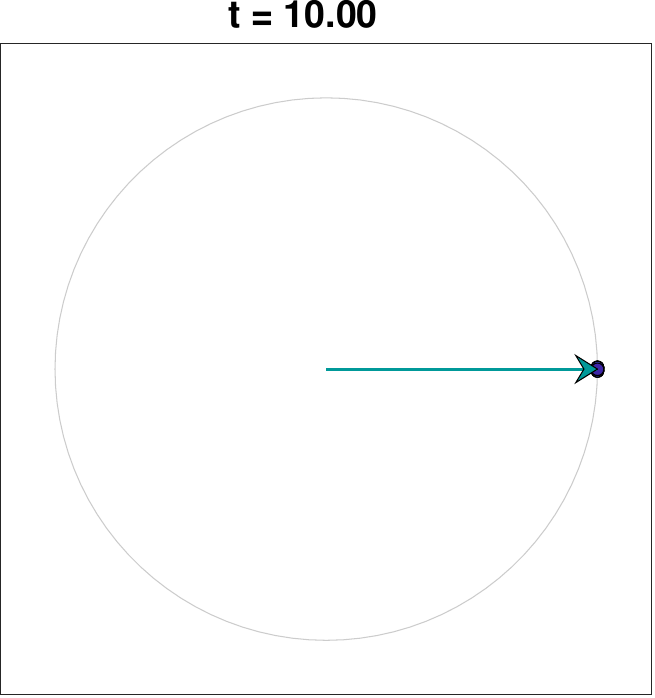}\qquad
            \includegraphics[scale=0.2]{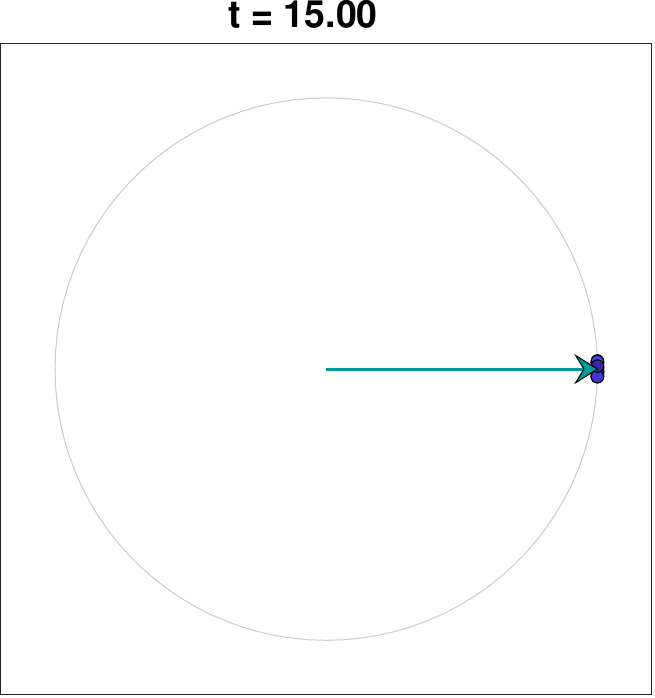}
        \end{minipage}
    }
    
     \subfloat[Pairwise phase differences histogram]{
    \begin{minipage}[c]{\textwidth}
            \centering
            \includegraphics[scale=0.2]{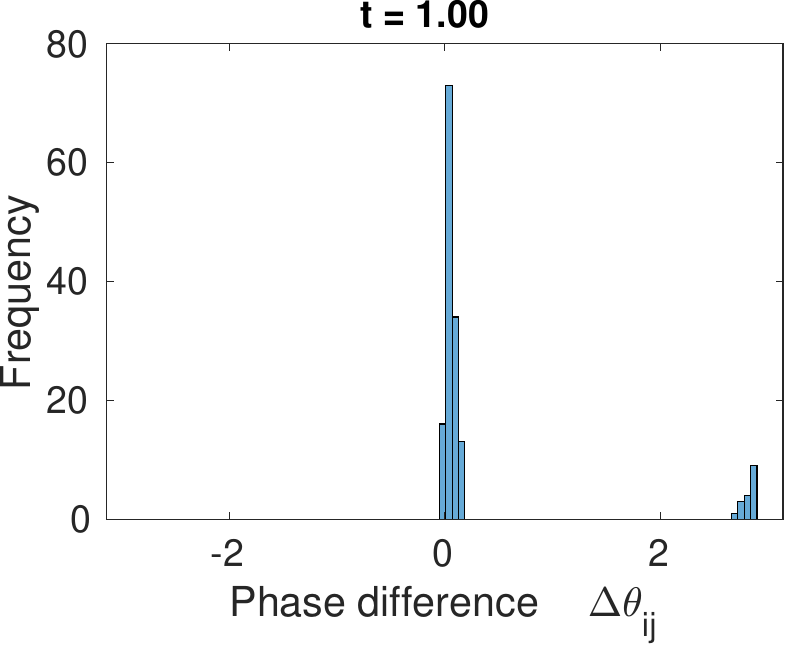}\quad
            \includegraphics[scale=0.2]{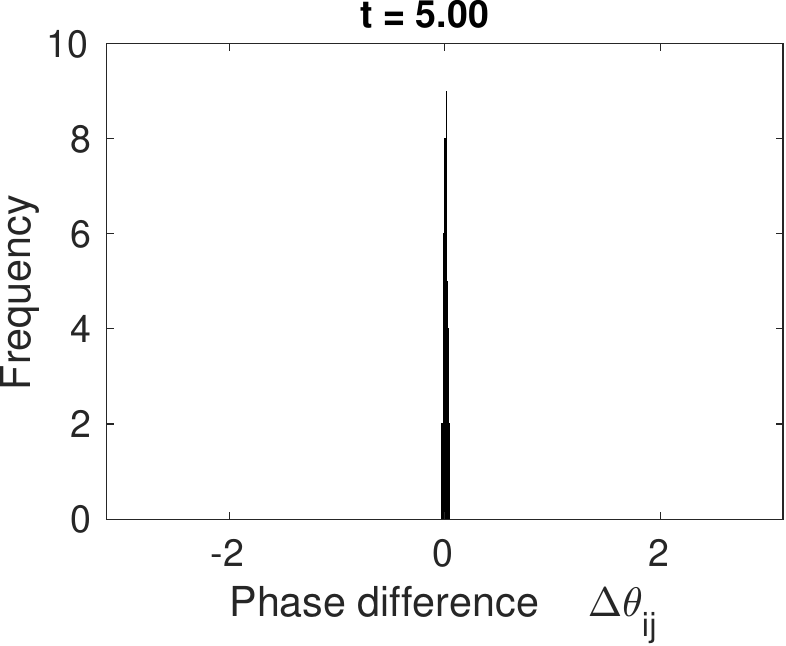}\quad
            \includegraphics[scale=0.2]{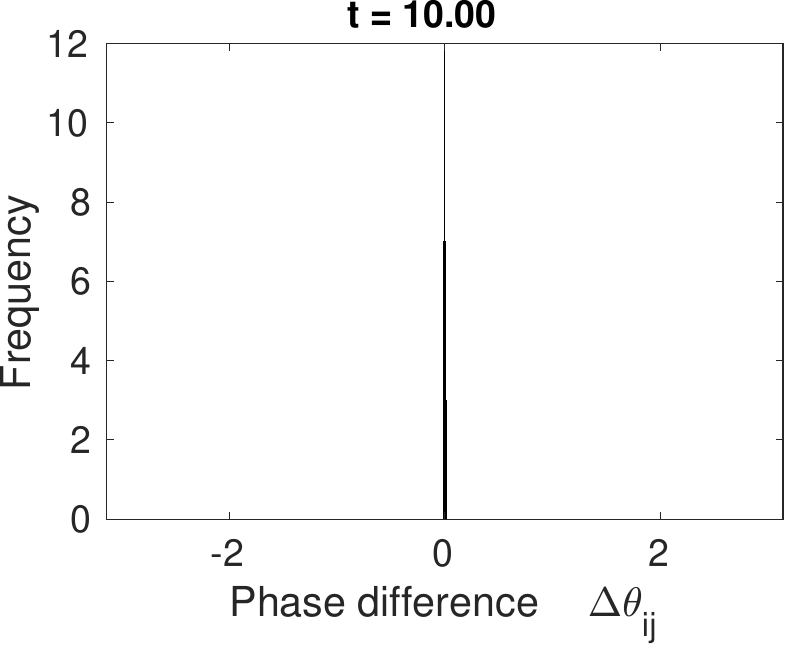}\quad
            \includegraphics[scale=0.2]{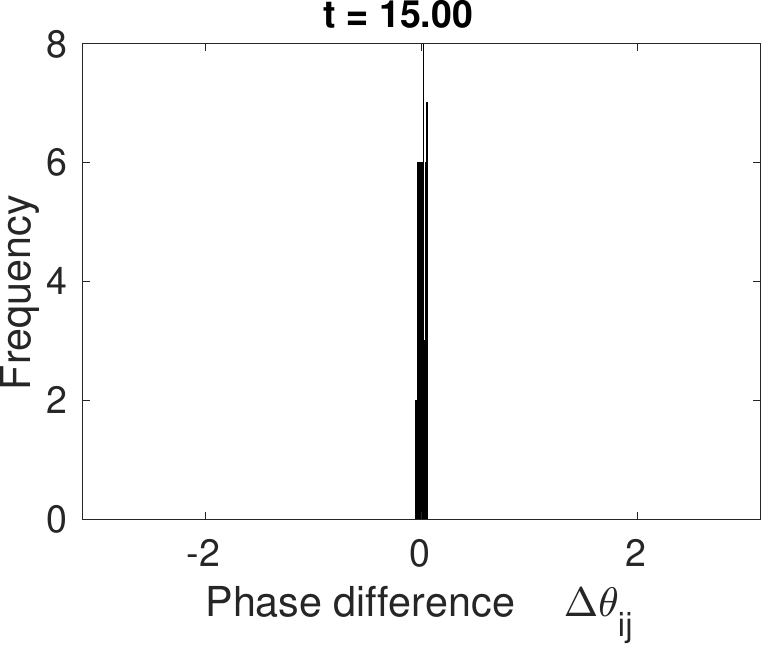}
        \end{minipage}
    }
    
    \subfloat[Pairwise phase differences heatmap]{
    \begin{minipage}[c]{\textwidth}
            \centering
            \includegraphics[scale=0.21]{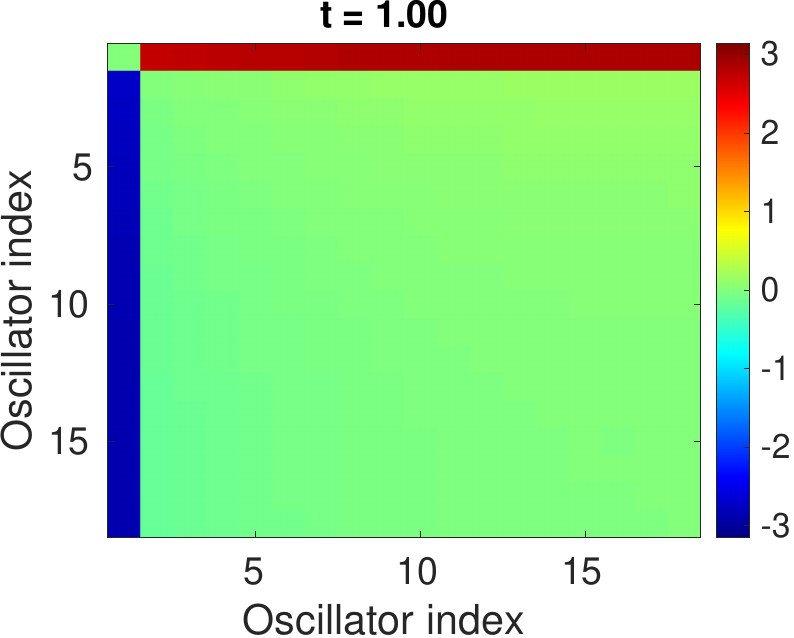}\quad
            \includegraphics[scale=0.21]{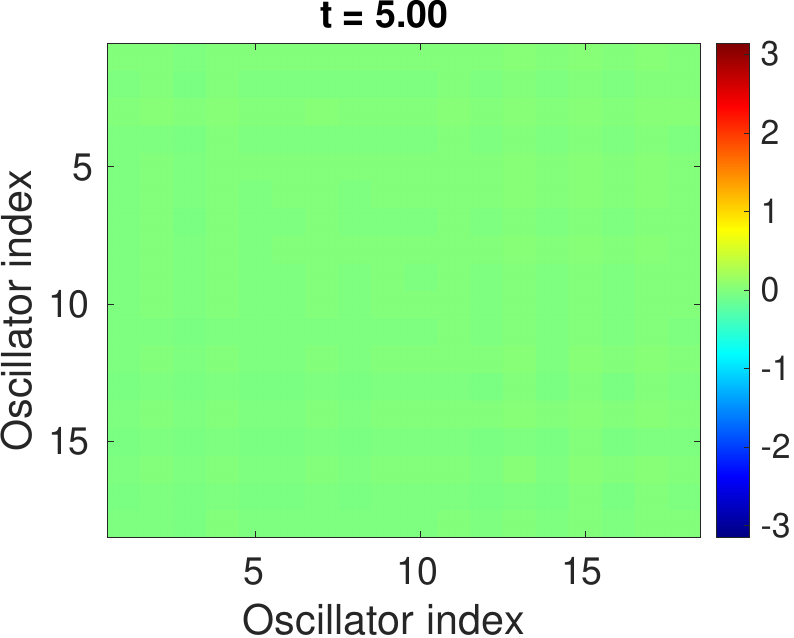}\quad
            \includegraphics[scale=0.21]{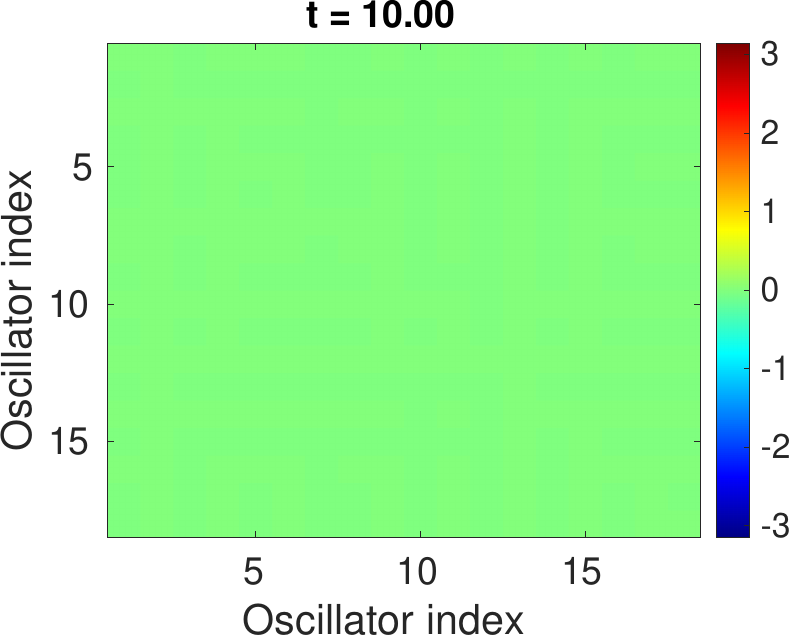}\quad
            \includegraphics[scale=0.21]{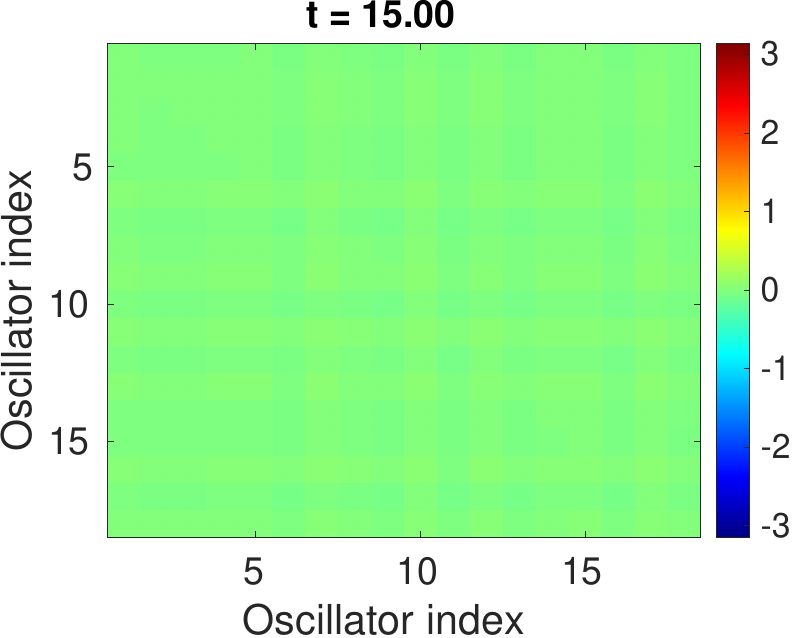}
        \end{minipage}
    }

    \caption{
    Visualization plots of the oscillator model show that, under a \emph{bidirectional, non-periodic, rendezvous protocol} communication (Figure~\ref{fig:bidir_topologies}), resynchronization among 18 GSSOR processes occurs in half the time compared to a unidirectional topology.
    }
    \label{fig:bidir-sync}
\end{figure*}

In MPI applications, knowing the state of each process at each moment in time allows us to calculate the phase difference $\theta_i - \theta_j$ between processes $i$ and $j$. 
It is important to note that in MPI, phase changes do not occur continuously but rather in discrete steps (iterations). 
For example, in a \emph{desynchronized} state, if at a given time processes $i$ and $j$ are two iterations apart, the phase difference would be $\theta_i - \theta_j = 2 \times 2\pi$. 
This phase information is subsequently utilized to generate the phase circle plots (traces displayed in subfigures (a) of Figures~\ref{fig:unidir-sync} through~\ref{fig:unidir-desync}, along with the corresponding phase circle plot videos referenced alongside), enabling direct qualitative comparison with those produced by the model.

Comparison with empirical MPI traces collected using Intel ITAC validates that the oscillator model captures several key phenomena observed in real-world MPI-parallel, bulk-synchronous and barrier-free applications. Several scenarios are covered in the following subsections. 

\subsubsection{Resynchronization in scalable, compute-bound workloads} \label{sec:sync}
The model captures the inherent tendency of a scalable, compute-bound GSSOR code to restore phase-aligned execution after perturbations, as demonstrated in Figure~\ref{fig:unidir-sync} \cite{AfzalHW19}. Among the various visualizations shown in (b)--(i), some are particularly well suited for capturing the process of resynchronization, i.e., the system's return to phase-aligned execution after perturbations subside. The order parameter plot in (b) is one of the most effective tools in this context, as it provides a clear and interpretable signal: A rising $R(t)$ indicates the restoration of global synchrony. Similarly, the potential plot in (f) effectively reveals system stabilization through a decreasing potential $V(t)$, capturing the convergence toward a lower-energy, synchronized state. The topological phase gradient in (d) offers additional insight by quantifying local phase disparities; a gradual decline in this metric signals improving alignment among neighboring oscillators. Supporting these, the synchronization entropy plot in (c) captures reductions in phase disorder and is particularly useful when partial or multimodal synchrony is involved. Pairwise phase difference timelines in (e) provide detailed oscillator-level resolution and visibly flatten as resynchronization occurs, though they may become visually complex for larger systems. In contrast, static visualizations such as the phase circle plot in (g), pairwise phase difference histogram in (h), and heatmap in (i) are best used for confirming the final synchronized state, rather than tracking the resynchronization process itself. Together, these tools offer complementary strengths, with dynamic plots being most effective for monitoring recovery over time, and static plots serving as snapshots of the different states.

  \smallskip \highlight{\emph{Upshot}:
  Time-resolved metrics like the order parameter, potential, and phase gradient are best suited for detecting the progression of resynchronization, while static visualizations confirm its outcome and reveal structural coherence at convergence.
}

\subsubsection{Topology-sensitive idle wave propagation in scalable, compute-bound workloads} \label{sec:topology}
Model simulations in Figure~\ref{fig:bidir-sync}(b-i) reproduce the topology-dependent resynchronization behavior observed in empirical ITAC traces (Figure~\ref{fig:bidir-sync}(a)), capturing how changing the communication pattern from uni-directional to bi-directional next-neighbor topology (using the rendezvous protocol in MPI) affects the absorption of idle waves in scalable MPI workloads, reduces resynchronization time by half \cite{AfzalHW2021}. In the \emph{order parameter plot} (b), $R(t)$ rises roughly twice as fast under bi-directional communication, indicating an earlier return to global synchrony. The \emph{potential plot} (f) similarly shows a faster decay in $V(t)$, reflecting a quicker drop in system energy. The \emph{topological phase gradient} (d) confirms this trend with steeper and earlier reductions in local phase misalignment, while the \emph{entropy plot} (c) shows a more rapid decline in disorder. The \emph{pairwise phase difference timelines} (e) offer a vivid contrast: trajectories flatten sooner and more uniformly in the bi-directional case. Static plots reinforce these dynamics at snapshot time points; the \emph{phase circle} (g) exhibits tighter clustering, the \emph{histogram} (h) shows a narrower peak, and the \emph{heatmap} (i) reveals more uniform pairwise alignment. Altogether, these metrics consistently demonstrate that bidirectional communication halves the time required for resynchronization, making the system markedly more efficient at recovering from perturbations.

  \smallskip \highlight{\emph{Upshot}:
  The oscillator model correctly maps the faster resynchronization of an idle wave when using the rendezvous protocol (which is modeled by a stronger coupling in the model). This is evident across both dynamic and static metrics within the oscillator model at various levels.
}

\begin{figure*}[hp]
    \centering
    \subfloat[Order parameter]{\includegraphics[scale=0.26]{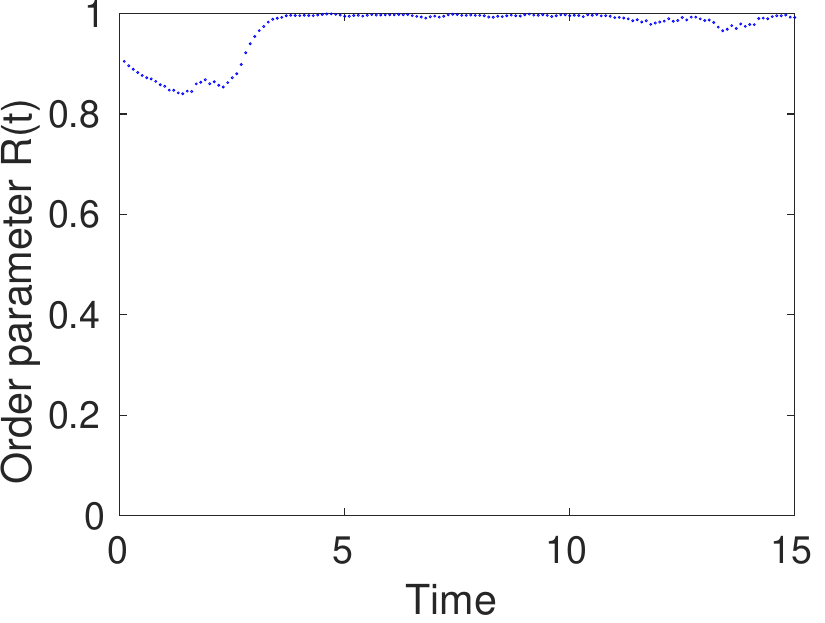}}\quad
    \subfloat[Entropy]{\includegraphics[scale=0.26]{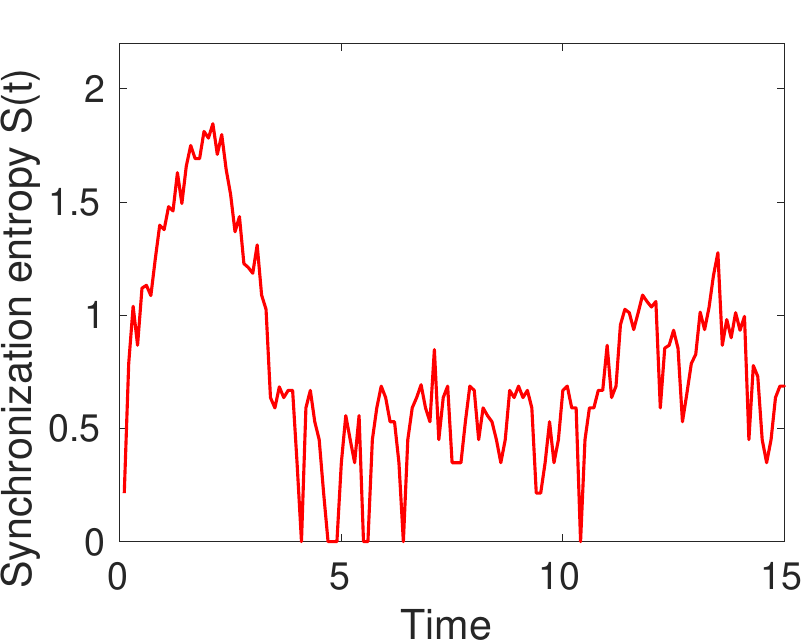}}\\[1ex]
    
    \subfloat[Phase gradient]{\includegraphics[scale=0.25]{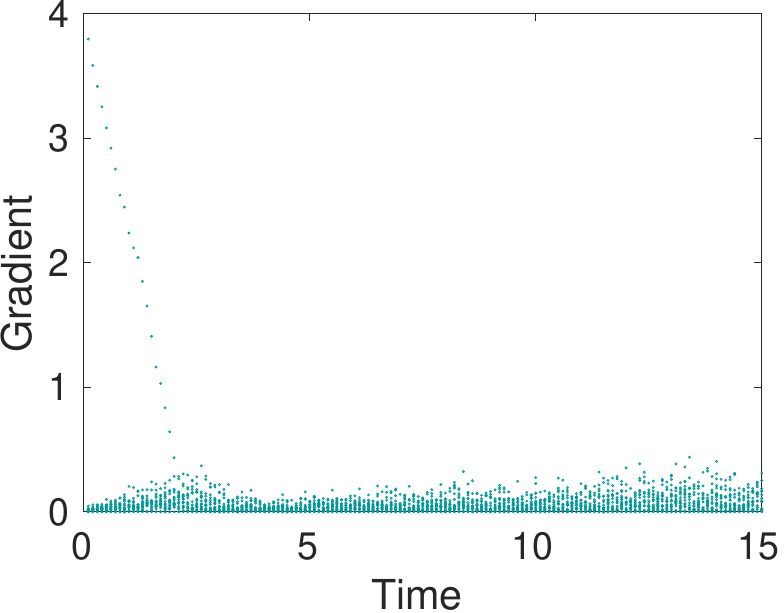}}\quad
    \subfloat[Phase differences]{\includegraphics[scale=0.25]{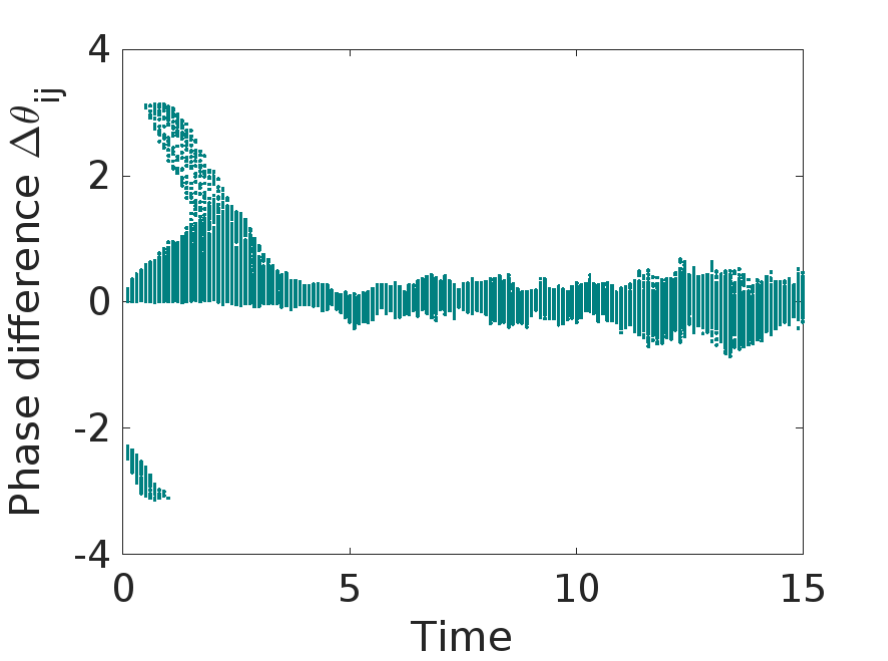}}\quad
    \subfloat[Potential]{\includegraphics[scale=0.3]{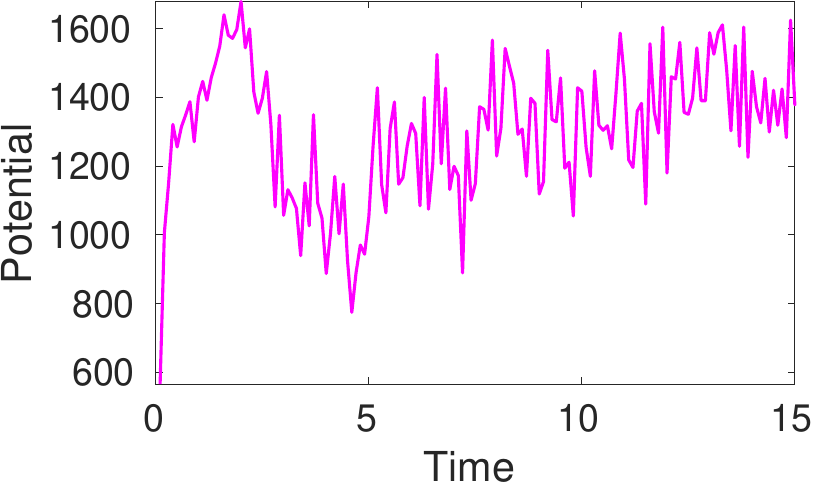}}\\[1ex]

     \subfloat[Phase circle]{
    \begin{minipage}[c]{\textwidth}
            \centering
            \includegraphics[scale=0.32]{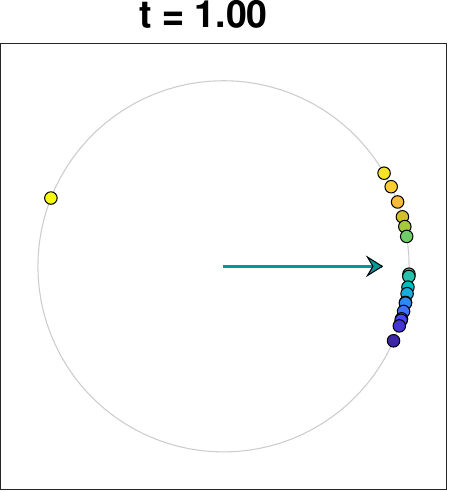}\qquad
            \includegraphics[scale=0.32]{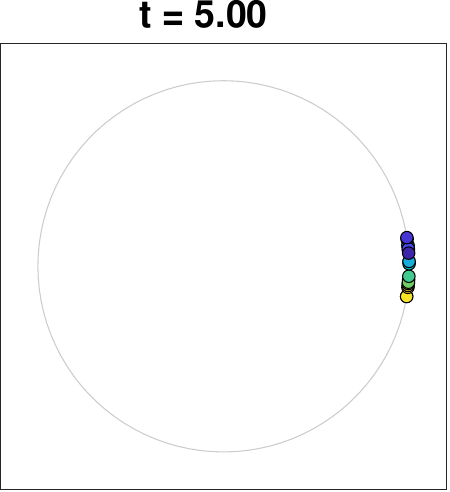}\qquad
            \includegraphics[scale=0.32]{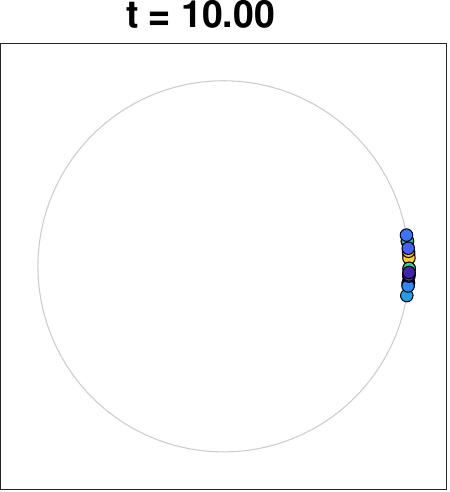}\qquad
            \includegraphics[scale=0.32]{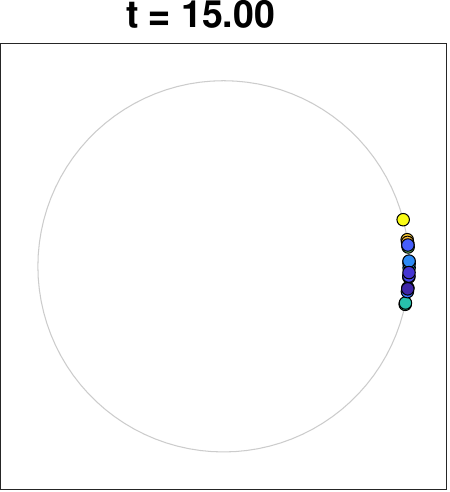}
        \end{minipage}
    }
    
     \subfloat[Pairwise phase differences histogram]{
    \begin{minipage}[c]{\textwidth}
            \centering
            \includegraphics[scale=0.22]{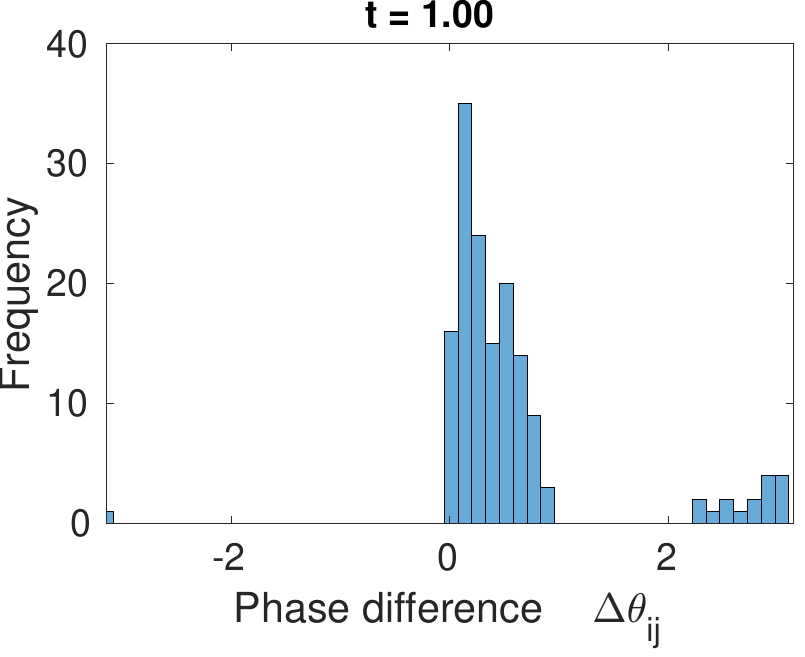}\quad
            \includegraphics[scale=0.22]{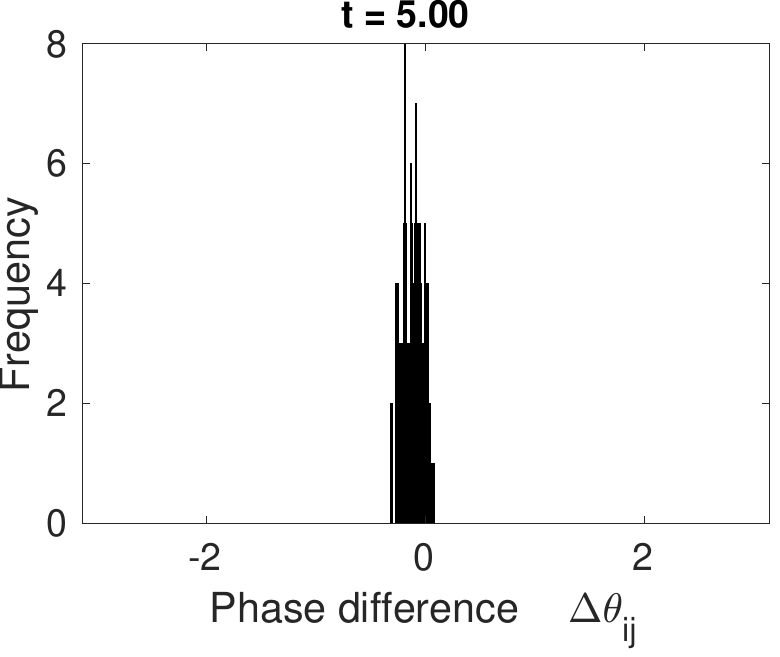}\quad
            \includegraphics[scale=0.22]{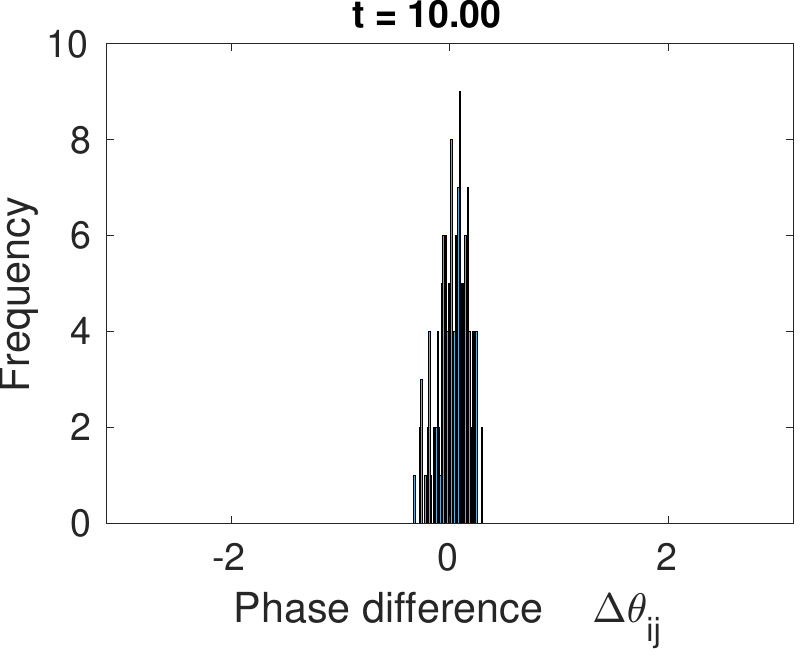}\quad
            \includegraphics[scale=0.22]{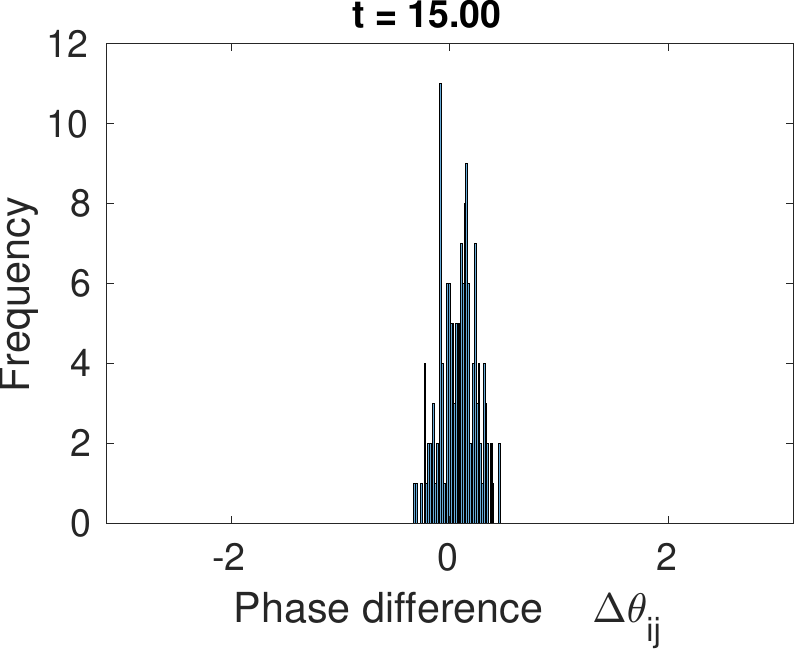}
        \end{minipage}
    }
    
    \subfloat[Pairwise phase differences heatmap]{
    \begin{minipage}[c]{\textwidth}
            \centering
            \includegraphics[scale=0.22]{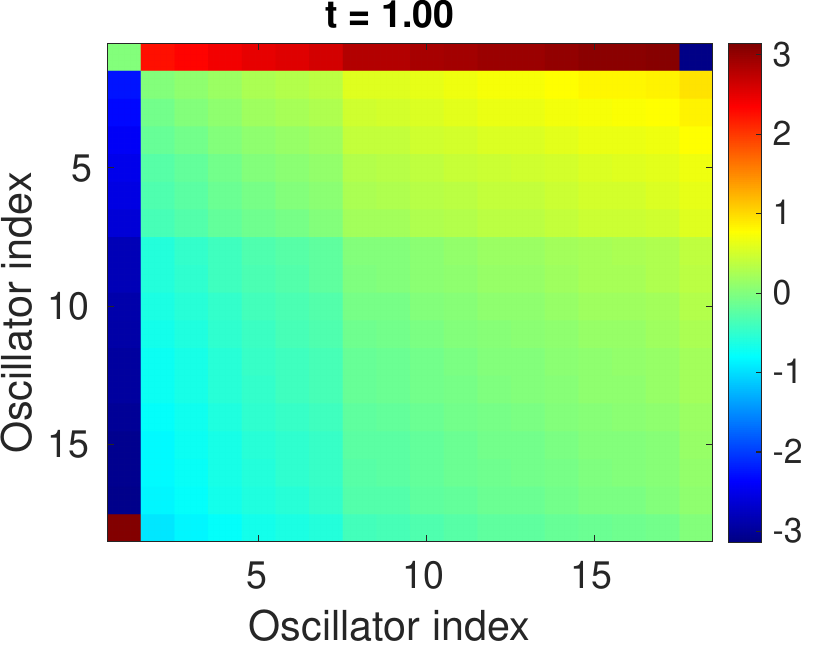}\quad
            \includegraphics[scale=0.22]{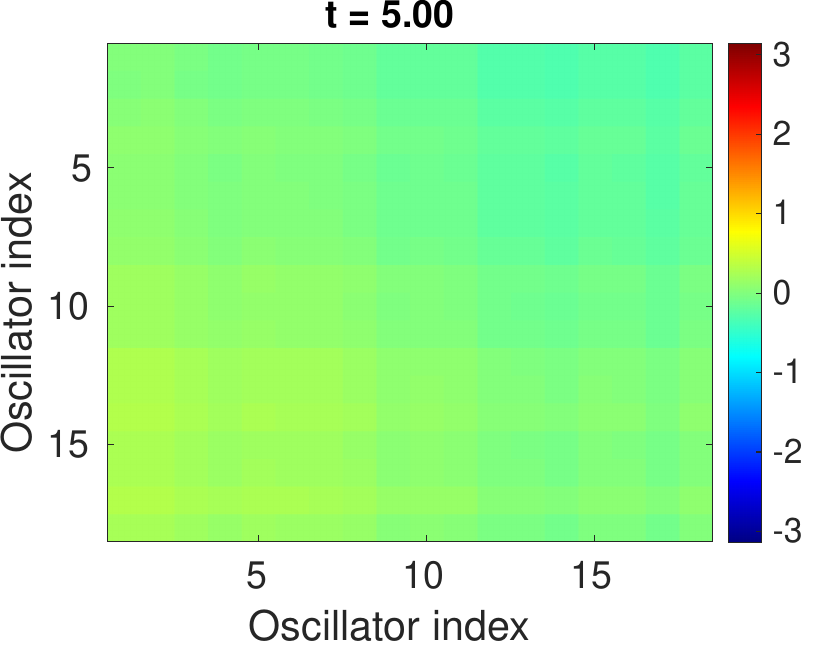}\quad
            \includegraphics[scale=0.22]{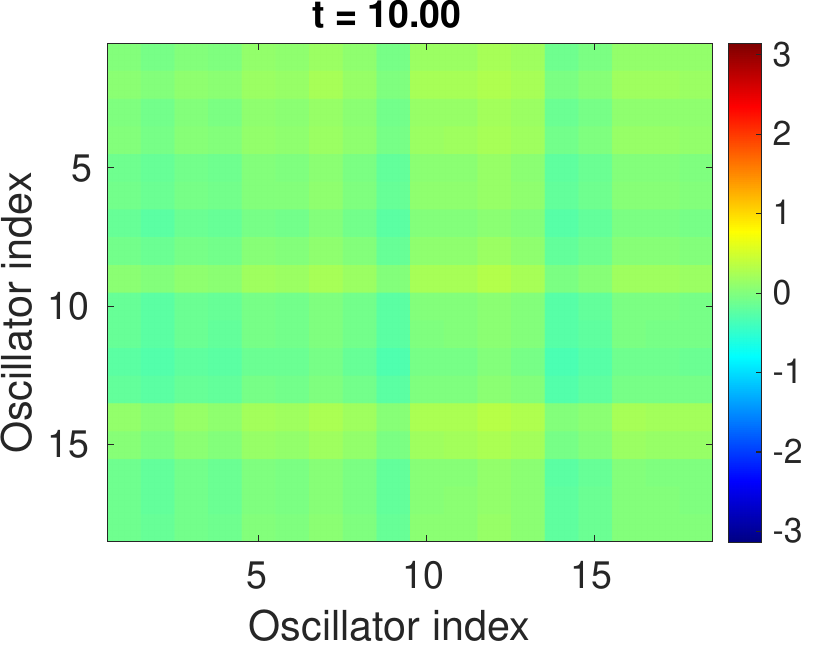}\quad
            \includegraphics[scale=0.22]{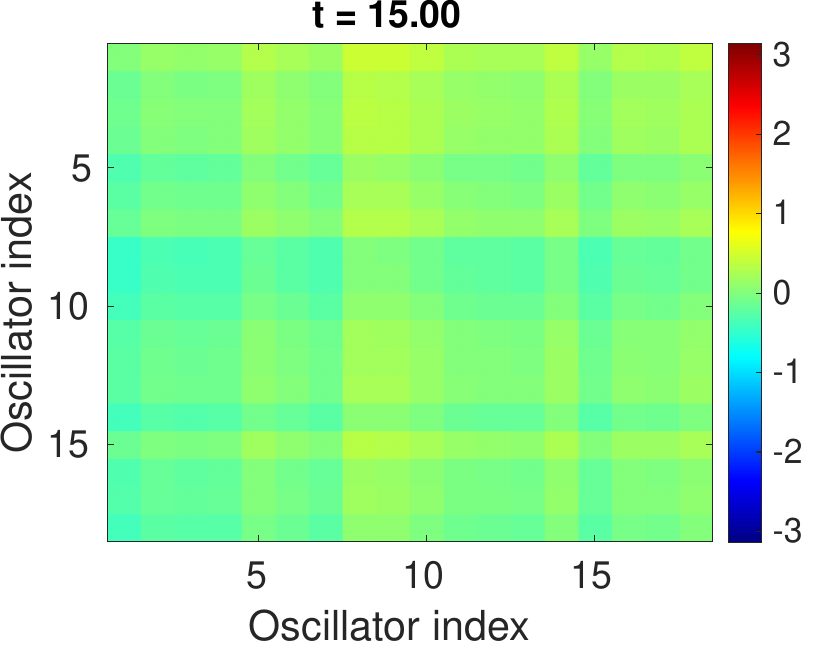}
        \end{minipage}
    }

    \caption{
        Visualization plots of the oscillator model demonstrate that, in the presence of \emph{local noise $\zeta_i(t) = 5\dot{\theta}_i(t) \cdot \mathbf{r}_i(t)$} (as described in Section~\ref{sec:testbed}), resynchronization among 18 GSSOR processes occurs rapidly local noise suppresses the propagating delay introduced in the initial condition.
    }
    \label{fig:unidir-sync-zeta1}
\end{figure*}

\begin{figure}[t]
    \centering
    \centering
    \subfloat[$\zeta_i(t) = 2\dot{\theta}_i(t) \cdot \mathbf{r}_i(t)$]{\includegraphics[width=0.3\textwidth]{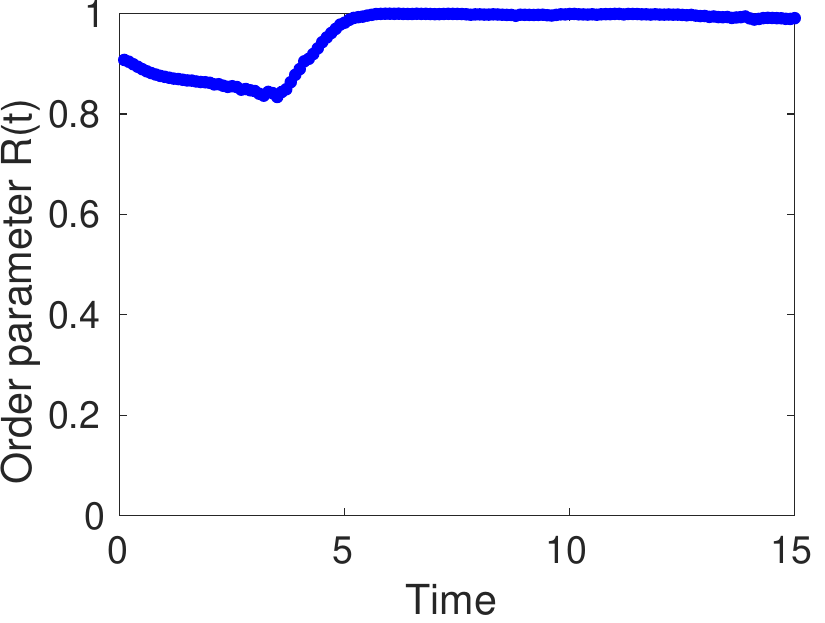}}\quad
    \subfloat[$\zeta_i(t) = 3\dot{\theta}_i(t) \cdot \mathbf{r}_i(t)$]{\includegraphics[width=0.3\textwidth]{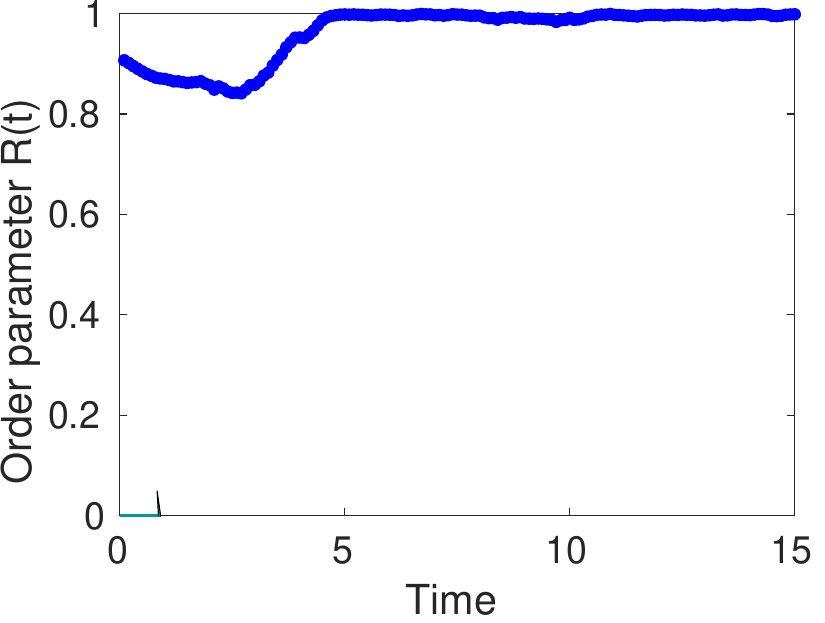}}\quad
    \subfloat[$\zeta_i(t) = 6\dot{\theta}_i(t) \cdot \mathbf{r}_i(t)$]{\includegraphics[width=0.3\textwidth]{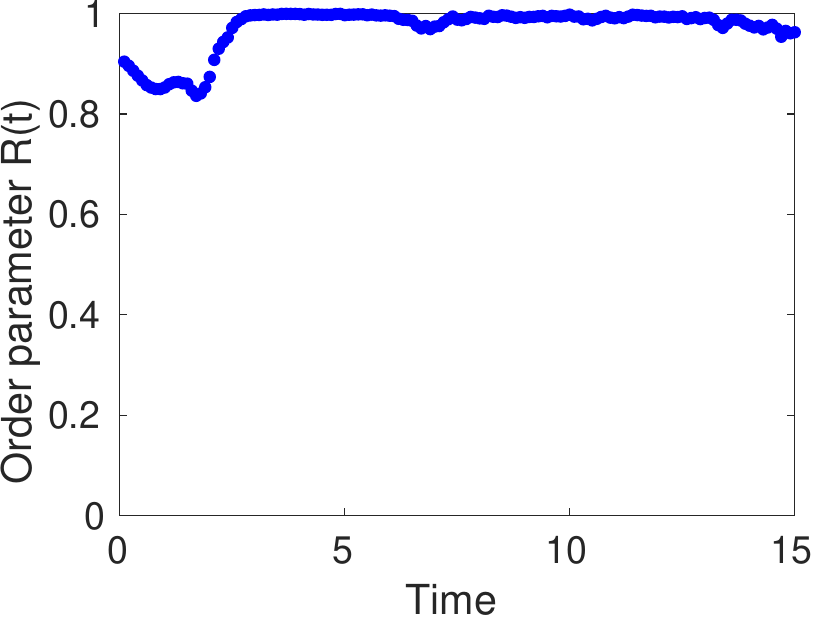}}\\[1ex]
    
    \subfloat[$\zeta_i(t) = 10\dot{\theta}_i(t) \cdot \mathbf{r}_i(t)$]{\includegraphics[width=0.3\textwidth]{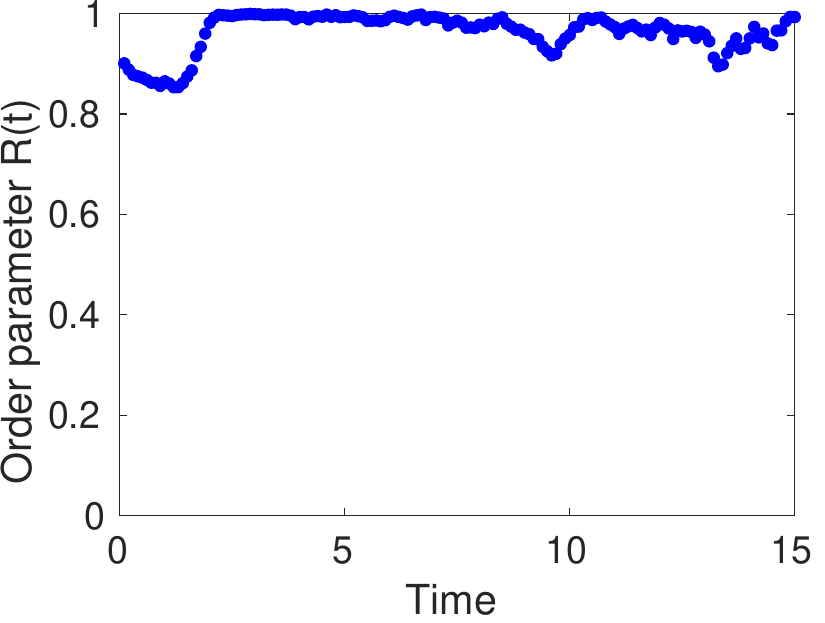}}\quad
    \subfloat[$\zeta_i(t) = 15\dot{\theta}_i(t) \cdot \mathbf{r}_i(t)$]{\includegraphics[width=0.3\textwidth]{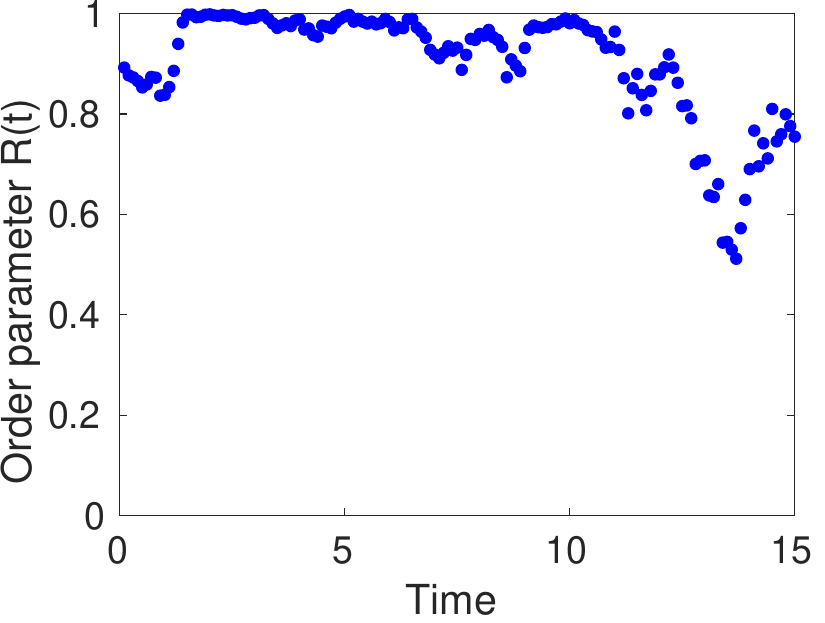}}\quad
    \subfloat[$\zeta_i(t) = 20 \dot{\theta}_i(t) \cdot \mathbf{r}_i(t)$]{\includegraphics[width=0.3\textwidth]{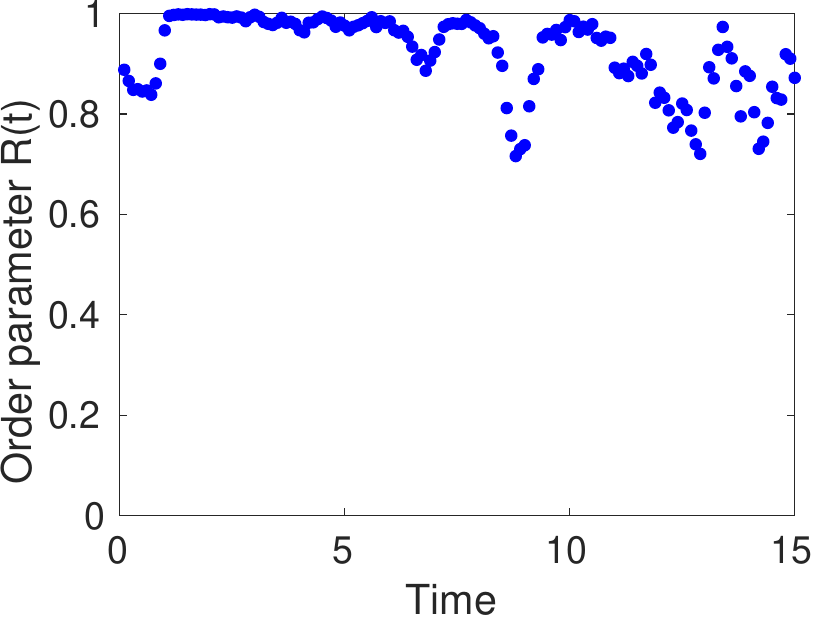}}\quad

    \caption{ Order parameter plots demonstrate rapid resynchronization across 18 processes with unidirectional, non-periodic communication as the $\zeta$ value increases from 2\% to 20\% of the current $\dot{\theta}_i(t)$ value. }
    \label{fig:unidir-sync-zeta2}
\end{figure}

\subsubsection{Noise-accelerated relaxation in scalable, compute-bound workloads} \label{sec:noise}
The model qualitatively corroborates that noise accelerates the relaxation of propagating delays in scalable, compute-bound workloads, a phenomenon also observed in trace-based evidence \cite{AfzalHW19}. As depicted in Figure~\ref{fig:unidir-sync-zeta1}(a), the presence of noise significantly speeds up the decay of phase misalignments, leading to quicker resynchronization and higher decay of propagating delays across the system. Among the various visualizations (b)--(i), several demonstrate this effect with particular clarity. In the \emph{order parameter plot} (b), the rising $R(t)$ curve is notably steeper under noisy conditions, indicating that global synchrony is achieved more rapidly. Similarly, the \emph{potential plot} (f) reveals a faster decay in the potential $V(t)$, suggesting a quicker transition to the lower-energy synchronized state. The \emph{topological phase gradient} plot (d) highlights how noise leads to a faster reduction in local phase misalignments, as evidenced by sharper drops in the phase gradient values. The \emph{entropy plot} (c) further illustrates the accelerating effect of noise, with a more pronounced and rapid decrease in synchronization disorder. The \emph{pairwise phase difference timelines} (e) reinforce this by showing faster convergence to synchronization across individual oscillator pairs. Finally, the static visualizations -- the \emph{phase circle} (g), the \emph{pairwise phase difference histogram} (h), and the \emph{heatmap} (i) -- show quicker and more consistent clustering in phase alignment, especially in the presence of noise. Together, these metrics confirm that noise accelerates the decay of propagating delays, leading to faster resynchronization and overall system stabilization in scalable, compute-bound MPI workloads.
Similarly, as depicted in Figure~\ref{fig:unidir-sync-zeta2}, increasing the $\zeta$ value (representing noise strength) leads to a faster rise in the order parameter $R(t)$, indicating quicker recovery of synchrony across the system. Higher $\zeta$ values, ranging from 1\% to 25\% of the current $\dot{\theta}_i(t)$ value, accelerate the resynchronization process.
 
\smallskip \highlight{\emph{Upshot}: 
Propagating delays in MPI decay more quickly under noise, leading to faster synchronization. Most metrics studied here are useful for visualizing this dependency. 
}

\begin{figure*}[hp]
    \centering
    \subfloat[MPI trace (\href{https://github.com/RRZE-HPC/OSC-AD/blob/main/MPI_trace_Jacobi/Jacobi_d-1.mp4}{video})]{\includegraphics[scale=0.42]{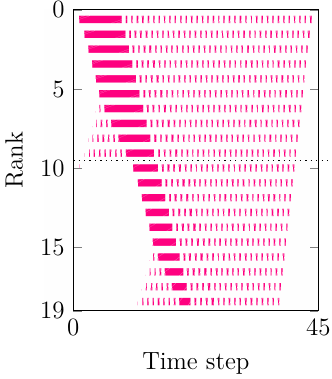}\quad\quad}
    \subfloat[Order parameter]{\includegraphics[scale=0.25]{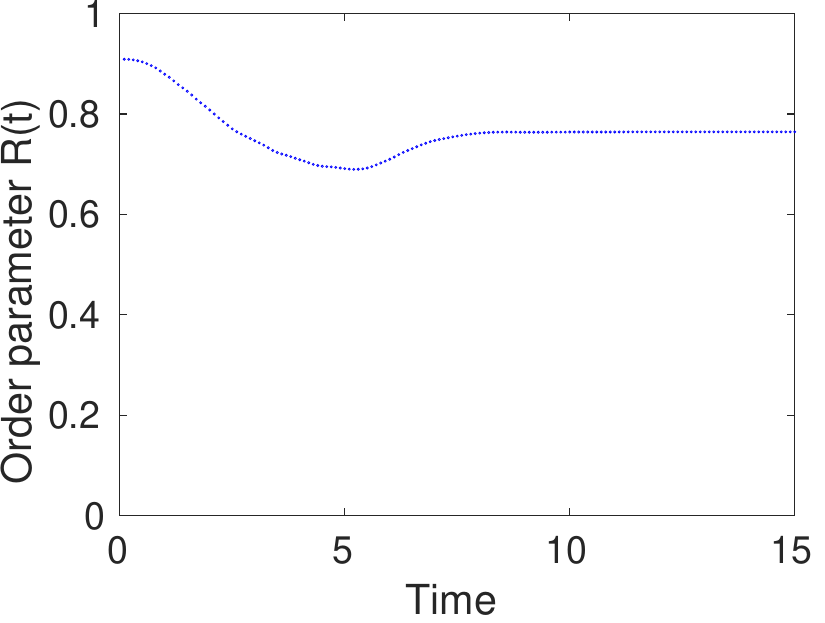}}\quad
    \subfloat[Entropy]{\includegraphics[scale=0.25]{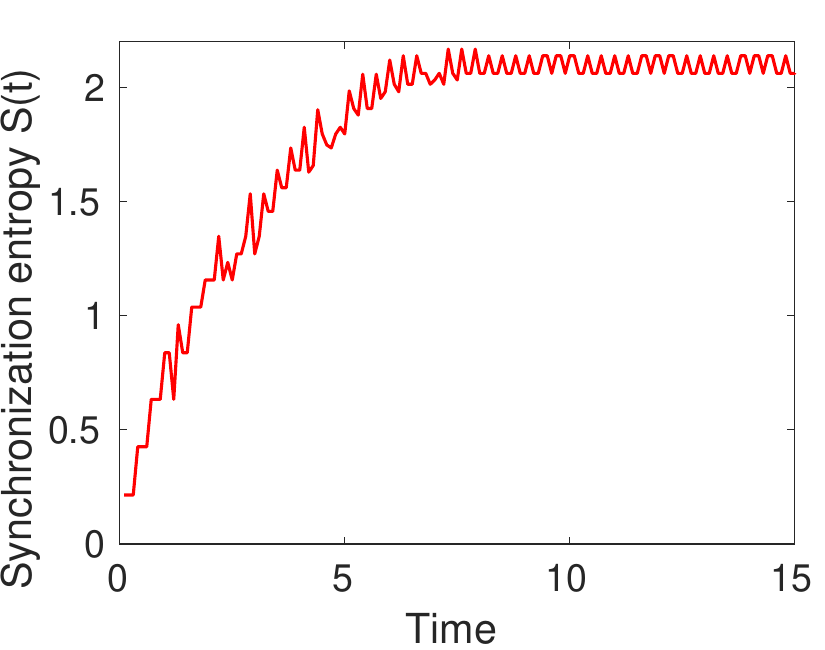}}\\[1ex]
    
    \subfloat[Phase gradient]{\includegraphics[scale=0.27]{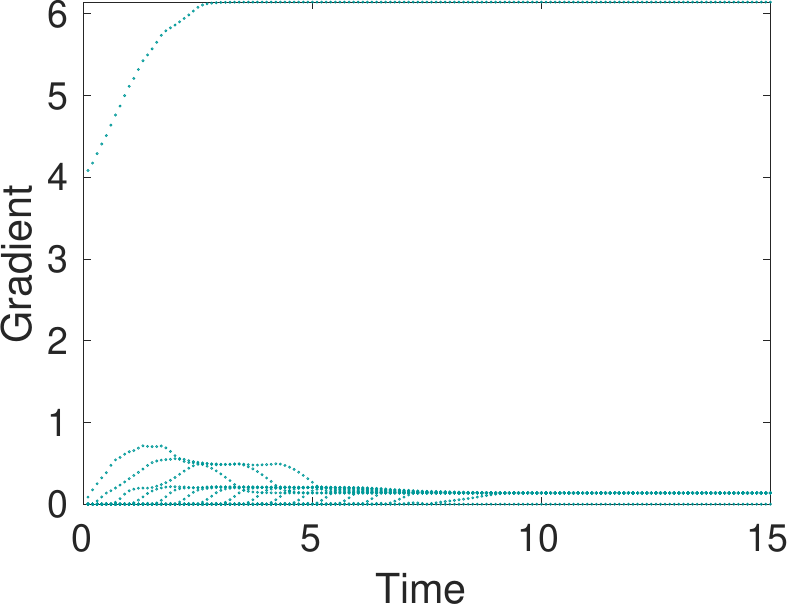}}\quad
    \subfloat[Phase differences]{\includegraphics[scale=0.27]{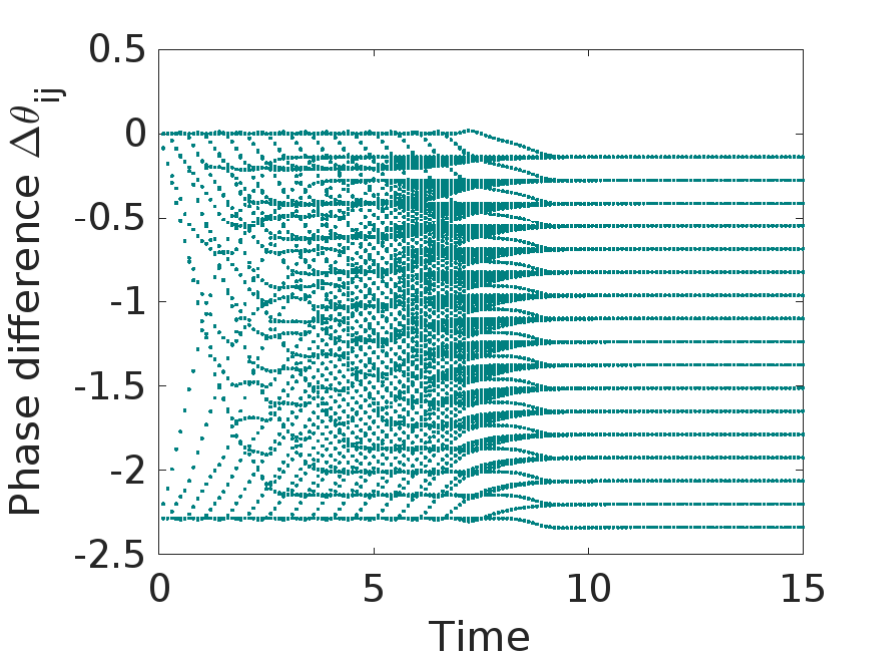}}\quad
    \subfloat[Potential]{\includegraphics[scale=0.27]{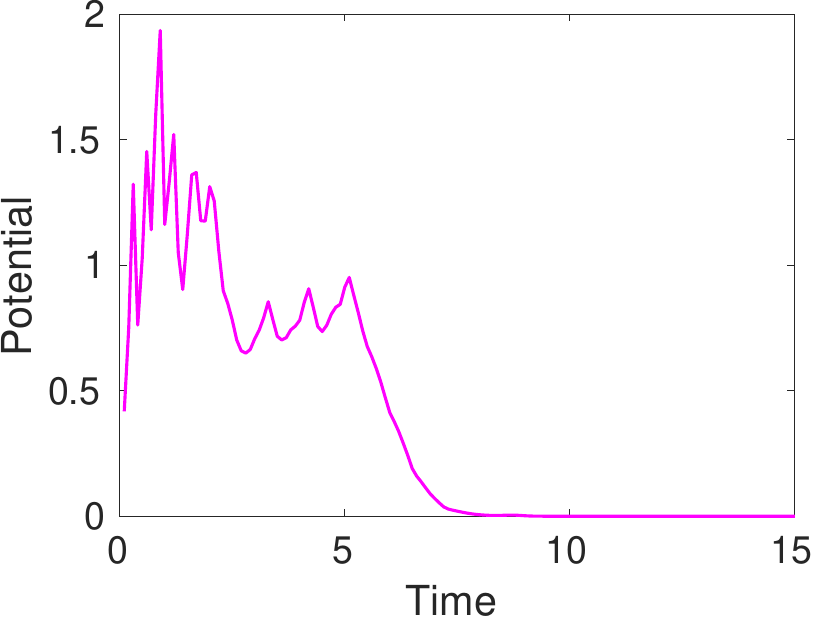}}\\[1ex]

     \subfloat[Phase circle]{
    \begin{minipage}[c]{\textwidth}
            \centering
            \includegraphics[scale=0.21]{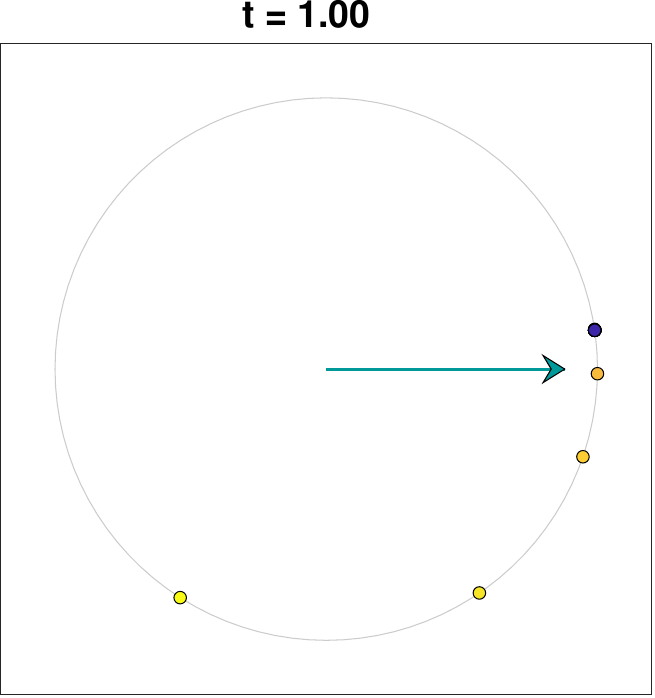}\qquad
            \includegraphics[scale=0.21]{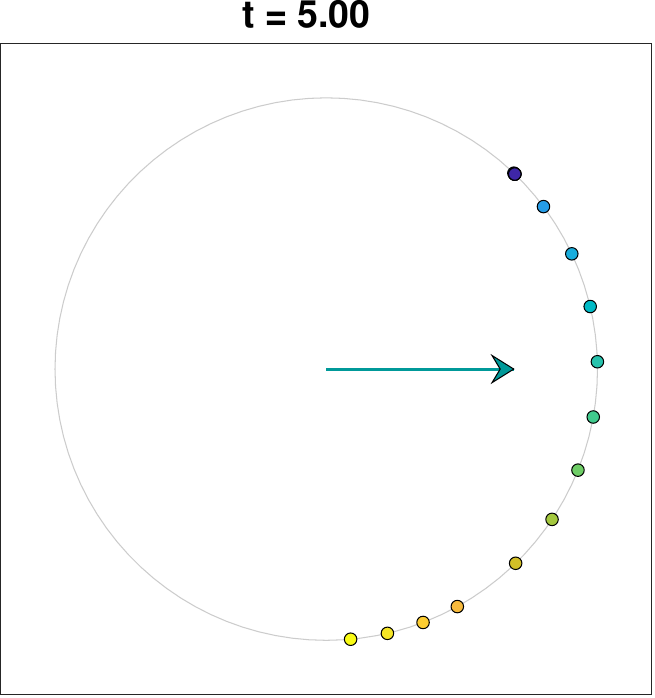}\qquad
            \includegraphics[scale=0.21]{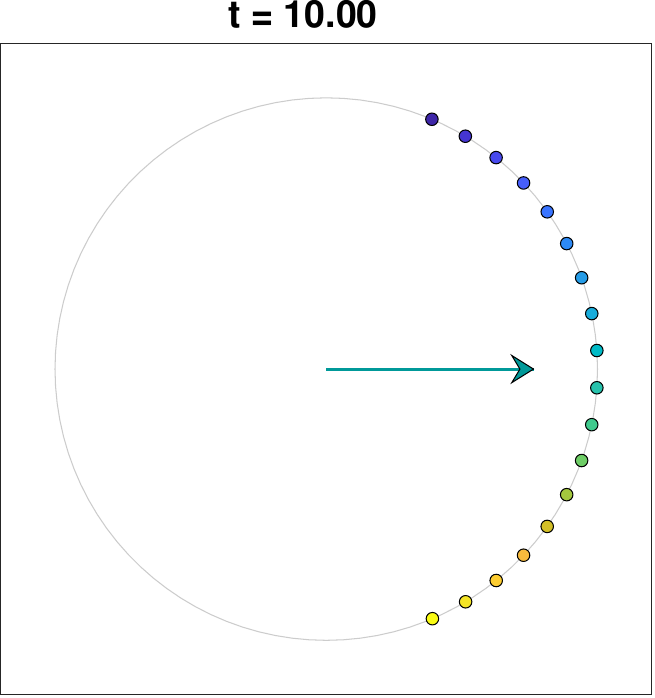}\qquad
            \includegraphics[scale=0.21]{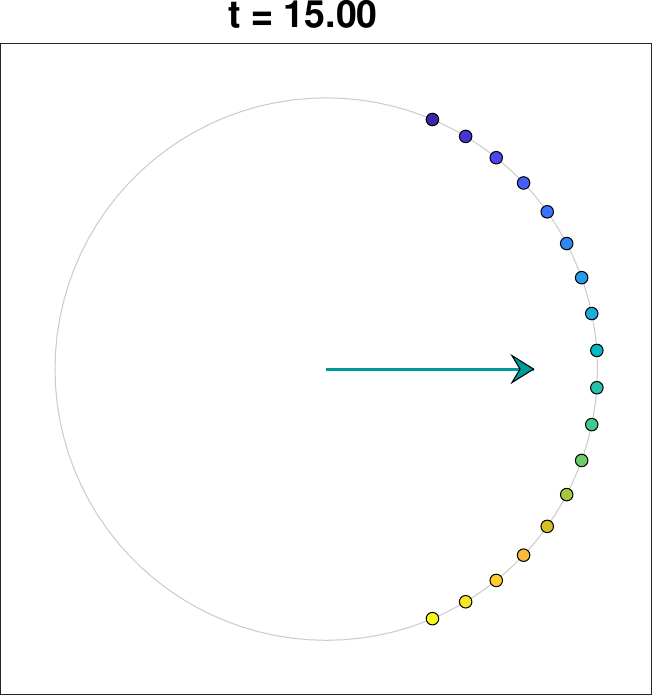}
        \end{minipage}
    }
    
     \subfloat[Pairwise phase differences histogram]{
    \begin{minipage}[c]{\textwidth}
            \centering
            \includegraphics[scale=0.21]{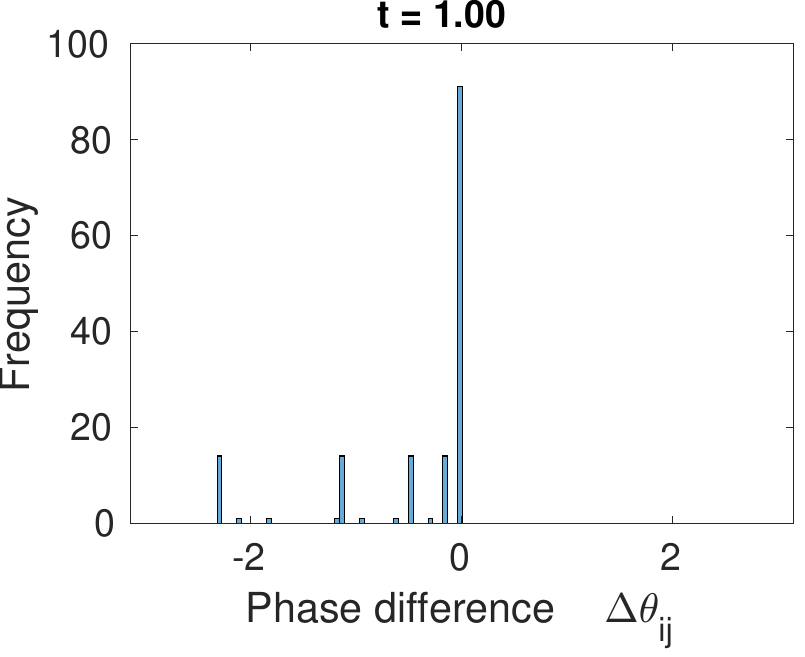}\quad
            \includegraphics[scale=0.21]{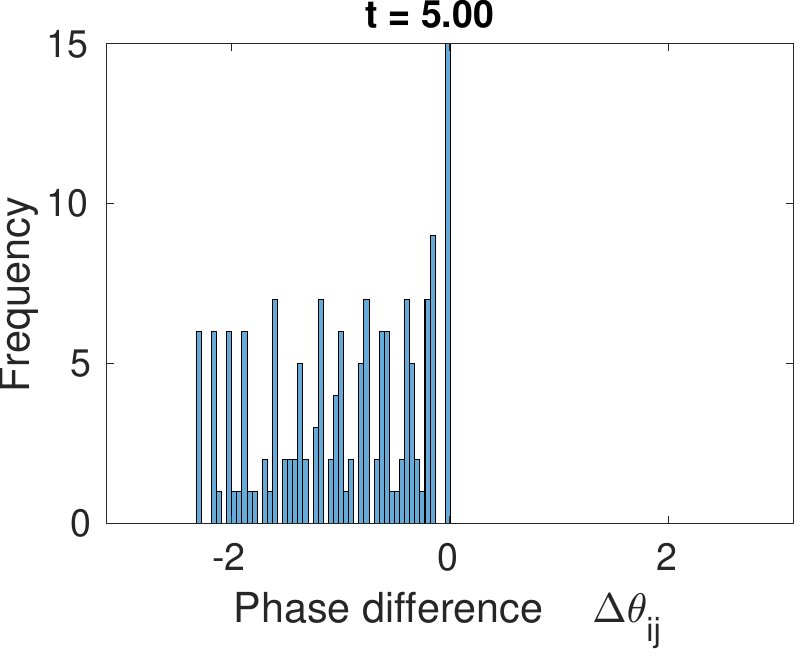}\quad
            \includegraphics[scale=0.21]{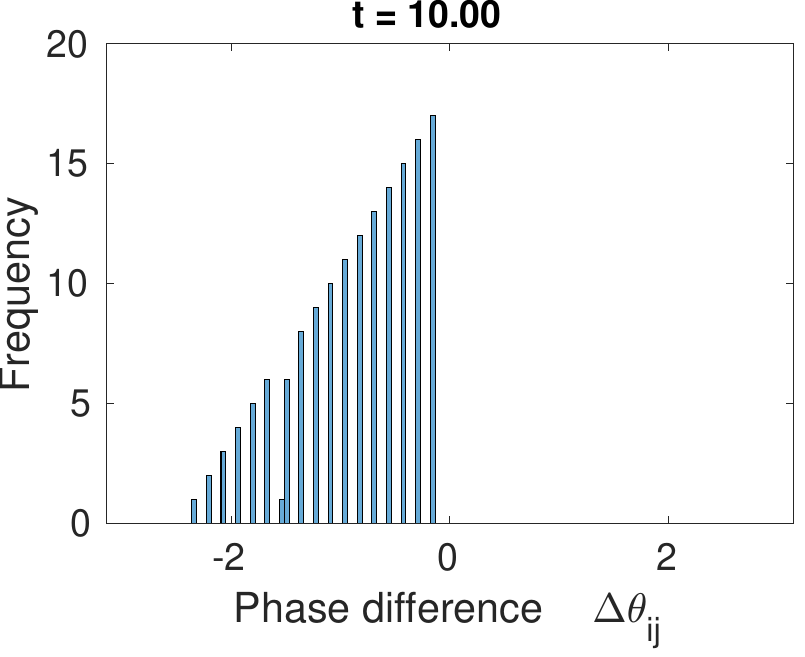}\quad
            \includegraphics[scale=0.21]{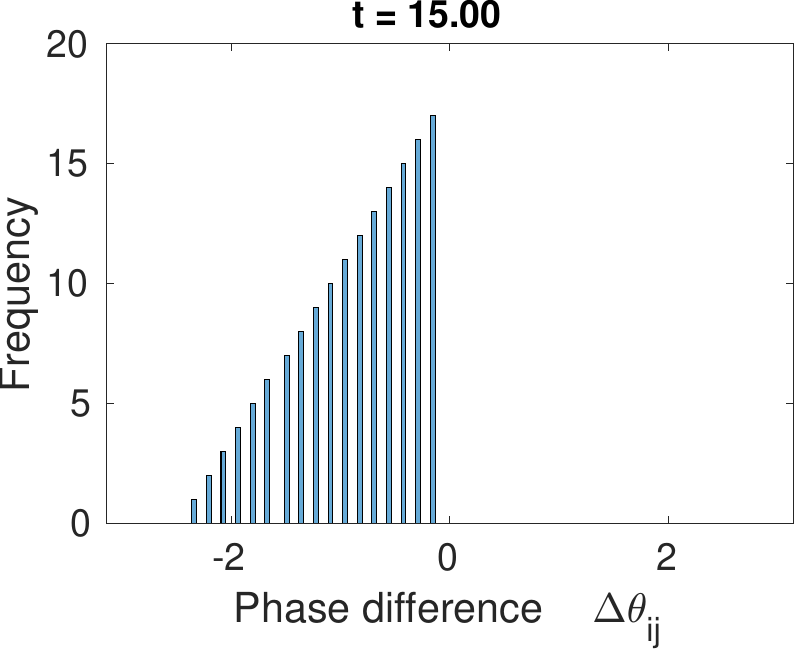}
        \end{minipage}
    }
    
    \subfloat[Pairwise phase differences heatmap]{
    \begin{minipage}[c]{\textwidth}
            \centering
            \includegraphics[scale=0.21]{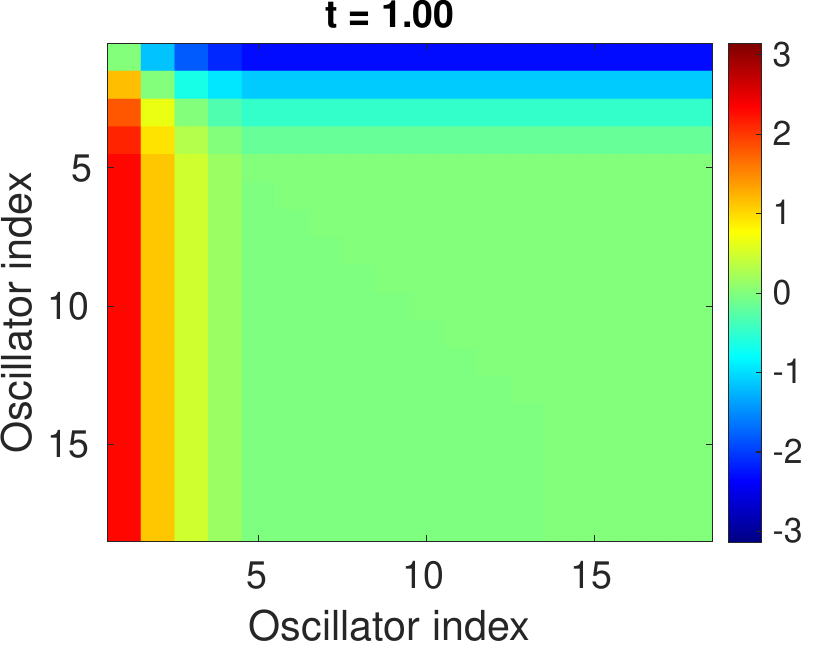}\quad
            \includegraphics[scale=0.21]{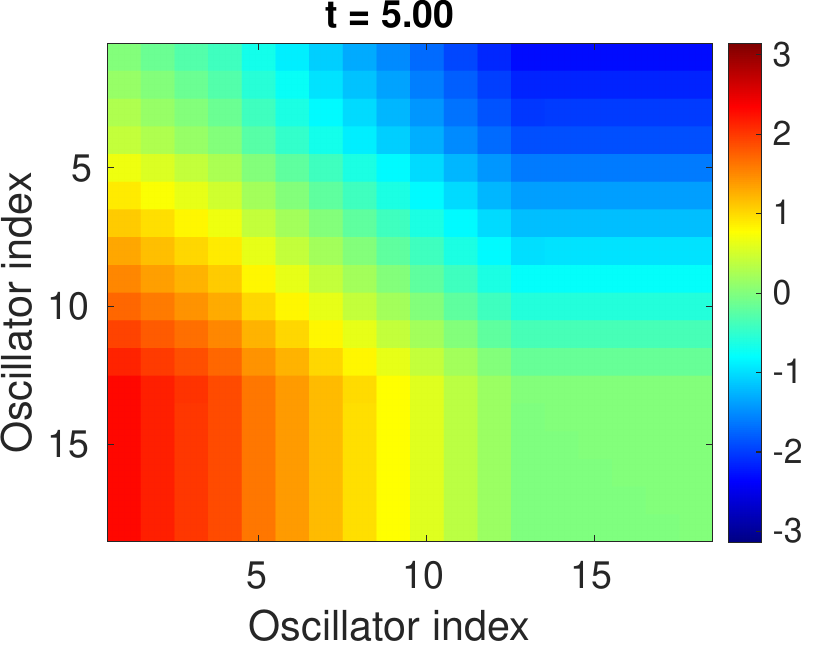}\quad
            \includegraphics[scale=0.21]{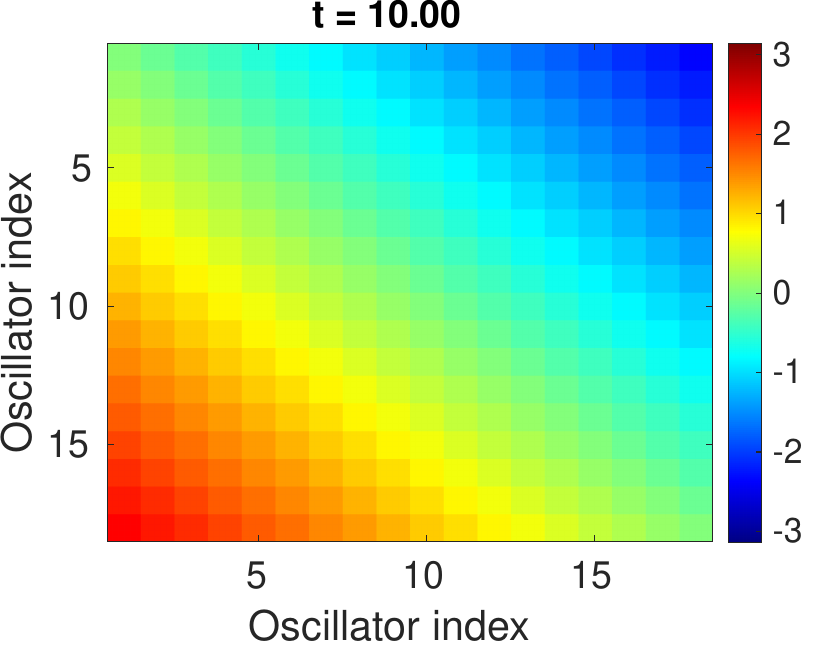}\quad
            \includegraphics[scale=0.21]{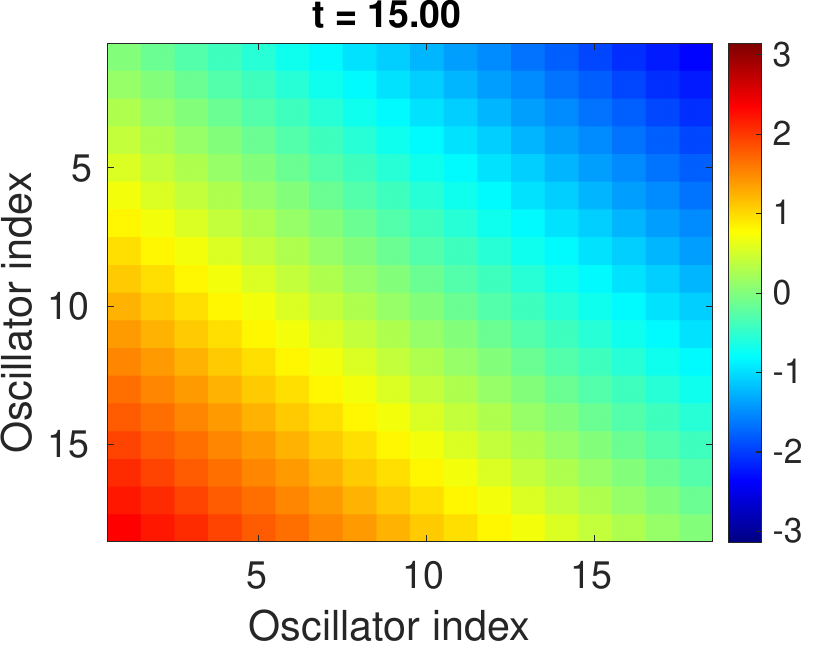}
        \end{minipage}
    }

    \caption{
        Visualization plots of the oscillator model depict structured desynchronization dynamics across 18 processes (oscillators) in the 2D 5-point Jacobi code following a delay induced on a single process. Communication is based on a \emph{unidirectional, non-periodic} topology.
    }
    \label{fig:unidir-desync}
\end{figure*}

\subsubsection{Desynchronization in bottlenecked, memory-bound workloads} \label{sec:desync}
As shown in Figure~\ref{fig:unidir-desync}, the model visualizations capture persistent phase drift characteristic of computational wavefronts observed in a memory-bound, bottlenecked 2D-5pt Jacobi smoother, closely matching ITAC trace visualizations. The system fails to restore phase alignment, with oscillators continuously drifting apart over time to avoid the bottleneck (see \cite{AfzalHW:2022:4} for a comprehensive discussion). Among the plots, the order parameter in (b) clearly illustrates \empty{sustained low} synchrony levels and the potential plot in (f) stabilizes near zero, indicating persistent desynchronization without recovery. The topological phase gradient in (d) remains elevated, signaling persistent local phase mismatches consistent with wavefront propagation. Synchronization entropy in (c) stays high throughout, confirming the sustained phase differences. Pairwise phase difference timelines in (e) reveal continuously diverging trajectories, visualizing the ongoing loss of alignment at the oscillator level. Static snapshots of the phase circle plot (g) display scattered phase distributions, while the pairwise phase difference histogram (h) and heatmap (i) exhibit broad, smeared distributions indicative of disordered phases. Together, these visualizations comprehensively demonstrate the inability of bottlenecked workloads to regain synchrony, highlighting the computational wavefront as a persistent desynchronization phenomenon. 

  \smallskip \highlight{\emph{Upshot}: 
  The model captures the inherent tendency of bottlenecked, memory-bound code to exhibit persistent phase drift (called computational wavefront) after perturbations. The metrics under consideration capture this behavior well, with phase difference heatmaps being particularly expressive. 
} \smallskip

\section{Conclusion and Outlook}\label{sec:conclusion}

\subsection{Summary of contributions}

We presented the progressive enhancement of an oscillator-based model~\cite{Afzal:2023:3} aimed at analyzing the collective temporal dynamics of MPI-parallel, bulk-synchronous and barrier-free applications. By conceptualizing each process as a phase-coupled oscillator, the model offers a complementary approach to conventional performance modeling, capturing fundamental phenomena inherent to MPI-parallel programs, including the topology-sensitive propagation and decay of delays, alongside emergent synchronization and desynchronization patterns. Through a combination of theoretical advancements and empirical validation using MPI traces collected from compute- and memory-bound workloads on an HPC cluster, we demonstrate that this oscillator-based approach provides a novel and physically interpretable perspective for understanding these critical dynamic behaviors.

First, to enable more realistic simulations that encompass a broader spectrum of performance phenomena observed in practice, the proposed model extends the classical Kuramoto framework by integrating key MPI scalability and bottleneck-related characteristics. These include directional and sparse communication topologies reflecting inter-process dependencies, time-varying interaction delays, load imbalances, heterogeneous coupling strengths, and customized interaction potentials designed to capture distinct scaling behaviors associated with hardware bottlenecks. Such behaviors include asymptotic regimes such as resynchronization (restoring translational symmetry) in compute-bound codes and persistent desynchronization, so called ``computational wavefronts~\cite{AfzalHW20,AfzalHW:2022:4},'' (broken translational symmetry) in memory-bound codes.
In particular, we refine the interaction potential for scalable parallel applications by advancing from the classical Kuramoto sinusoidal function $\sin(\theta)$ to a scaled hyperbolic tangent function $\tanh(s\theta)$ with a slope parameter $s$, allowing different progression characteristics towards the synchronized state. 

Second, the evaluation methodology has been substantially extended to support comprehensive validation. We introduce and systematically assess a diverse set of analytical metrics and visualizations, spanning scalar, structural, and distribution-based perspectives, ranging from the classical phase circle view to order parameters, synchronization entropy, local coupling gradients, energy landscape representations, and both temporal and statistical views of pairwise phase differences (timeline, histogram, and heatmap). This spectrum of metrics may serve as an analytical toolkit for quantifying global synchronization, desynchronization, and localized coupling dynamics in MPI-parallel, bulk-synchronous, barrier-free applications, enabling systematic exploration of system responses to controlled perturbations, initial phase configurations, communication topologies, and workload-induced variability.

We think that these enhancements position the oscillator-based model as more than a mere conceptual analogy to MPI-parallel programs. Together with a more rigorous analytical treatment, it may well evolve to become a lightweight, physically grounded, and interpretable method for reasoning about performance dynamics.

\subsection{Directions for future research}

While the presented oscillator-based model advances the physical interpretation of collective temporal dynamics in MPI-parallel applications, several avenues for future exploration remain open.

\paragraph{Model refinement and extension} 
Future efforts should focus on tighter integration with real-time MPI traces (possibly via trace analysis tools) and automated topology extraction to enhance model realism, using the proposed curated collection of plots for guidance. Potential extensions include support for MPI programs with qualitatively distinct phases in terms of execution and communication (leading to time-dependent coupling constants and natural frequencies in the model); support for asynchronous, task-based, and hybrid MPI+OpenMP programming models; adaptation to heterogeneous HPC architectures (including GPUs and specialized accelerators) and the handling of hierarchical system topologies.
These enhancements will improve the model's applicability across diverse workloads.

\paragraph{Analytical and theoretical investigations} 
Pursuing formal analysis in the limit $P \to \infty$ could yield continuum or PDE approximations, providing insights into stability, wave propagation, and critical synchronization phenomena. This may uncover computational analogs to symmetry-breaking effects, enriching theoretical understanding. Further exploration of such phenomena may lead to the discovery of analogs to Goldstone modes or phase transitions in computational settings.

\paragraph{Hardware-software co-design and optimization} 
Modeling hardware characteristics such as NUMA effects, network topology asymmetry, and congestion can guide the co-design of HPC interconnects and scheduling policies. Investigating oscillator-derived performance metrics, such as phase variance minimization and energy dissipation rates, offers promising optimization strategies. Furthermore, examining how synchronization patterns evolve under hardware faults, transient errors, or system noise could inform the development of resilient and fault-tolerant scheduling strategies for large-scale HPC environments.

Collectively, by refining the model's scope and deepening its analytical treatment, the proposed oscillator-based framework could evolve into a versatile and predictive tool for HPC performance modeling. It holds the potential to bridge the gap between empirical trace analysis and a first-principles understanding of the complex dynamics in large-scale parallel computing systems.


\bibliographystyle{elsarticle-num-names} 
\bibliography{references}

\end{document}